\documentclass[prd,preprintnumbers,twocolumn,amsmath,nofootinbib,amssymb]{revtex4}
\usepackage{graphicx,color,dcolumn,booktabs,bm}
\usepackage{longtable,lscape}
\usepackage{txfonts}
\usepackage{overpic}
\usepackage{amssymb}
\usepackage{epstopdf}
\usepackage{indentfirst}
\usepackage{feynmf}   
\usepackage{slashed}  
\usepackage{cases}
\usepackage{color}
\usepackage{float}
\usepackage{multirow}
\usepackage{ulem}
\usepackage{graphicx,color,dcolumn,booktabs,bm}
\usepackage{epsfig,dsfont,amssymb,amsmath,amsfonts,amsbsy,mathrsfs}

\graphicspath{{Figures/}} %

\usepackage{hyperref}
\hypersetup{colorlinks,citecolor=blue,anchorcolor=red,menucolor=red, linkcolor=red,filecolor=red,runcolor=red,urlcolor=blue,frenchlinks=red}

\makeatletter
\@addtoreset{equation}{section}
\makeatother

\allowdisplaybreaks

\begin{document}

\title{Exotic double-charm molecular states with hidden or open strangeness and around $4.5\sim 4.7$ GeV}

\author{Fu-Lai Wang$^{1,2}$}
\email{wangfl2016@lzu.edu.cn}
\author{Xiang Liu$^{1,2}$\footnote{Corresponding author}}
\email{xiangliu@lzu.edu.cn}
\affiliation{$^1$School of Physical Science and Technology, Lanzhou University, Lanzhou 730000, China\\
$^2$Research Center for Hadron and CSR Physics, Lanzhou University and Institute of Modern Physics of CAS, Lanzhou 730000, China}

\begin{abstract}
In this work, we investigate the interactions between the charmed-strange meson  ($D_s, D_s^{*}$) in $H$-doublet and the (anti-)charmed-strange meson  ($D_{s1}, D_{s2}^{*}$) in $T$-doublet, where the one boson exchange model is adopted by considering the $S$-$D$ wave mixing and the coupled-channel effects. By extracting the effective potentials for the discussed $H_s\bar{T}_s$ and $H_s{T}_s$ systems, we try to find the bound state solutions for the corresponding systems. We predict the possible hidden-charm hadronic molecular states with hidden strangeness, i.e., the $D_s^{*}\bar D_{s1}+c.c.$ states with $J^{PC}$=$0^{--}, 0^{-+}$ and the $D_s^{*}\bar D_{s2}^{*}+c.c.$ states with $J^{PC}$=$1^{--}, 1^{-+}$. Applying the same theoretical framework, we also discuss the $H_s T_s$ systems. Unfortunately,
the existence of the open-charm and open-strange molecular states corresponding to the $H_s T_s$ systems can be excluded.
\end{abstract}

\maketitle

\section{Introduction}\label{sec1}

Studying exotic hadronic states, which are very different from conventional mesons and baryons, is an  intriguing research frontier full of opportunities and challenges
 in hadron physics. As an important part of hadron spectroscopy, exotic state can be as a good platform for deepening our understanding of non-perturbative behavior of quantum chromodynamics (QCD). Since the observation of the charmonium-like state $X(3872)$ \cite{Choi:2003ue}, a series of $X/Y/Z/P_c$ states have been observed in the past 17 years, which stimulated extensive discussions on exotic hadronic state assignments to them. Because of the masses of several $X/Y/Z/P_c$ states are close to the thresholds of two hadrons, it is natural to consider them as the candidates of the hadronic molecules, which is the reason why exploring hadronic molecular states has become popular. The theoretical and experimental progress on the hidden-charm multiquark states can be found by Refs. \cite{Chen:2016qju,Liu:2013waa,Hosaka:2016pey,Liu:2019zoy,Brambilla:2019esw,Olsen:2017bmm,Guo:2017jvc}. Among them, a big surprise is the observation from the  LHCb Collaboration of three $P_c$ states ($P_c(4312)$, $P_c(4440)$, and $P_c(4457)$) in 2019 \cite{Aaij:2019vzc}, which provides strong evidence to support these $P_c$ states as the $\Sigma_c \bar D^{(*)}$-type hidden-charm pentaquark molecules \cite{Li:2014gra,Karliner:2015ina,Wu:2010jy,Wang:2011rga,Yang:2011wz,Wu:2012md,Chen:2015loa}.

 Before presenting our motivation, we firstly need to give a brief review of how these observed $XYZ$ states were decoded as the corresponding hidden-charm molecular states.
 In 2003, $X(3872)$ was discovered by Belle \cite{Choi:2003ue}. For solving its low mass puzzle, the $D\bar{D}^*$ molecular state explanation was proposed in Ref. \cite{Wong:2003xk}. Later, more theoretical groups joined the discussion of whether $X(3872)$ can be assigned as $D\bar{D}^*$ molecular state \cite{Swanson:2003tb,Suzuki:2005ha}, where more and more effects were added in realistic calculation of the $D\bar{D}^*$ interaction \cite{Liu:2008fh,Thomas:2008ja,Liu:2008tn,Lee:2009hy,Zhao:2014gqa,Li:2012cs}. At the same time, another special $XYZ$ state is $Z^+(4430)$, which was observed by Belle \cite{Choi:2007wga}. In Refs. \cite{Liu:2007bf,Liu:2008xz,Close:2009ag}, a dynamic calculation of $Z^+(4430)$ as $D^*\bar D_1^{(\prime)}$ molecular state was performed. Later, the observed $Y(4140)$ \cite{Aaltonen:2009tz} also stimulated a universal molecular state explanation to $Y(3930)$ and $Y(4140)$ \cite{Liu:2009ei}, which is due to the similarity between $Y(3930)$ \cite{Uehara:2005qd} and $Y(4140)$ \cite{Aaltonen:2009tz}. Additionally, $Y(4260)$ \cite{Aubert:2005rm} as a $D\bar D_1$ molecular state was given\cite{Ding:2008gr} and discussed \cite{Cleven:2013mka,Wang:2013kra}. Besides $Y(4140)$ \cite{Aaltonen:2009tz}, the above studies on these typical charmonium-like $XYZ$ are mainly involved in hidden-charm hadronic molecular states without strangeness. These observed $XYZ$ states also result in several systematic theoretical calculations of the interaction between charmed and anti-charmed mesons \cite{Sun:2012zzd,Sun:2012sy,Hu:2010fg,Shen:2010ky,Chen:2015add}. {Since the $Y(4274)$ \cite{Aaij:2016nsc} is observed in the $J/\psi \phi$ invariant mass spectrum and is just below the threshold of the $D_s\bar{D}_{s0}(2317)$ channel, the $Y(4274)$ fits well to be the $S$-wave $D_s\bar{D}_{s0}(2317)$ molecular state with $J^P=0^-$ in Ref. \cite{Liu:2010hf}}. Obviously, these studies enlarged our knowledge of hidden-charm hadronic molecular states with mass below 4.5 GeV.

In the past years, more $XYZ$ states with higher mass were announced by many experiment collaborations \cite{Aaij:2016iza,Jia:2019gfe,Pakhlova:2008vn,Abazov:2007sf}.
In 2019, the BESIII Collaboration announced a white paper on the future physics program \cite{Ablikim:2019hff}, where they plan to perform a detailed scan of cross sections between 4.0 and 4.6 GeV and take more data above 4.6 GeV. These new measurements will not only result in complement the higher radial and orbital charmonium family \cite{Wang:2019mhs,Wang:2020prx}, but also provide us a good opportunity to study exotic state assignments to the $XYZ$ states above 4.5 GeV.

For hidden-charm molecular states with and without strangeness, which has mass below 4.5 GeV, we have abundant theoretical study. However, our knowledge of hidden-charm molecular states above 4.5 GeV is still not enough. Considering this research status and future experimental plan, we propose to explore exotic double-charm molecular states with hidden or open strangeness existing in mass range around $4.5\sim 4.7$ GeV, which are relevant to the interactions between $S$-wave charmed-strange meson in $H$-doublet and $P$-wave (anti-)charmed-strange meson in $T$-doublet. Generally, the $H_s\bar{T}_s$ system corresponds to hidden-charm and hidden-strange hadronic molecular state with the $(c\bar{s})(\bar c {s})$ configuration while the $H_s{T}_s$ system is involved in open-charm and open-strange hadronic molecular state with the $(c\bar{s})(c\bar{s})$ configuration.
In the following, the $H_s\bar{T}_s$ and $H_s{T}_s$ systems will be main body of this work.

For obtaining the interaction information of the $H_s\bar{T}_s$ and $H_s{T}_s$  systems, we apply one boson exchange (OBE) model \cite{Tornqvist:1993ng,Tornqvist:1993vu} to deduce the effective potentials in coordinate space. With this effective potential reflecting the interaction of the $H_s\bar{T}_s$ and $H_s{T}_s$  systems, we try to find bound state solutions of these discussed  $H_s\bar{T}_s$ and $H_s{T}_s$ systems, by which we may predict possible exotic double-charm molecular states with hidden or open strangeness around $4.5\sim 4.7$ GeV mass range. Further suggestion of experimental search for this new type of hadronic molecular states will be given.

This paper is organized as the follows. After introduction, we illustrate the deduction of the effective potentials of  these discussed  $H_s\bar{T}_s$ and $H_s{T}_s$ systems in Sec. \ref{sec2}. With this preparation, we present the numerical results of finding the bound state solutions in Sec. \ref{sec3}. Finally, this work ends with a short summary in Sec. \ref{sec4}.

\section{Effective potentials involved in the $H_s\bar{T}_s$ and $H_s{T}_s$ systems}\label{sec2}

In this section, we deduce the effective potentials in the coordinate space for the $H_s\bar{T}_s$ and $H_s{T}_s$ systems, where the OBE model is adopted in concrete calculation.
Here, we need to emphasize that the OBE model was extensively applied to study these observed $X/Y/Z/P_c$ states \cite{Chen:2016qju,Liu:2019zoy}.

\subsection{Effective Lagrangians}\label{seb21}

When describing the interactions quantitatively at the hadronic level, we use the effective Lagrangian approach. For writing out the compact effective Lagrangians related to the charmed meson in $H$-doublet and the (anti-)charmed meson in $T$-doublet, it is convenient to introduce the super-fields $H^{(Q)}_a$, $T^{(Q)\mu}_a$, $H^{(\overline{Q})}_a$, $T^{(\overline{Q})\mu}_a$, and their corresponding conjugate fields \cite{Ding:2008gr}. According to the heavy quark limit \cite{Wise:1992hn}, the super-fields $H^{(Q)}_a$ and $T^{(Q)\mu}_a$ corresponding to the heavy-light meson $Q\bar{q}$ can be defined by \cite{Ding:2008gr}
\begin{eqnarray}
H^{(Q)}_a&=&{\cal P}_{+}\left(D^{*(Q)\mu}_a\gamma_{\mu}-D^{(Q)}_a\gamma_5\right),\nonumber\\
T^{(Q)\mu}_a&=&{\cal P}_{+}\left[D^{*(Q)\mu\nu}_{2a}\gamma_{\nu}-\sqrt{\frac{3}{2}}D^{(Q)}_{1a\nu}\gamma_5\left(g^{\mu\nu}-\frac{1}{3}\gamma^{\nu}\left(\gamma^{\mu}-v^{\mu}\right)\right)\right],\nonumber
\end{eqnarray}
respectively. Meanwhile, the anti-meson $\bar{Q}q$ super-fields $H^{(\overline{Q})}_a$ and $T^{(\overline{Q})\mu}_a$ can be obtained by the charge conjugate transformation for the super-fields $H^{(Q)}_a$ and $T^{(Q)\mu}_{a}$ \cite{Ding:2008gr}, where its expression denotes
\begin{eqnarray}
H^{(\overline{Q})}_a&=&\left(\bar{D}^{*(\overline{Q})\mu}_{a}\gamma_{\mu}-\bar{D}^{(\overline{Q})}_a\gamma_5\right){\cal P}_{-},\nonumber\\
T^{(\overline{Q})\mu}_{a}&=&\left[\bar{D}^{*(\overline{Q})\mu\nu}_{2a}\gamma_{\nu}-\sqrt{\frac{3}{2}}\bar{D}^{(\overline{Q})}_{1a\nu}\gamma_5\left(g^{\mu\nu}-\frac{1}{3}\left(\gamma^{\mu}-v^{\mu}\right)\gamma^{\nu}\right)\right]{\cal P}_{-}.\nonumber
\end{eqnarray}
Here, ${\cal P}_{\pm}=(1\pm{v}\!\!\!\slash)/2$ denotes the projection operator, and $v^{\mu}=(1,\,0,\,0,\,0)$ is the four velocity under the non-relativistic approximation. Besides, their conjugate fields can be expressed as
\begin{equation}
\overline{X}=\gamma_0X^{\dagger}\gamma_0,~~~X=H^{(Q)}_a,\,T^{(Q)\mu}_a,\,H^{(\overline{Q})}_a,\,T^{(\overline{Q})\mu}_{a}.
\end{equation}

On the basis of the heavy quark symmetry, the chiral symmetry, and the hidden local symmetry \cite{Wise:1992hn,Casalbuoni:1992gi,Casalbuoni:1996pg,Yan:1992gz}, the compact effective Lagrangians depicting the interactions between the (anti-)charmed mesons and light pseudoscalar and vector mesons were constructed in Ref. \cite{Ding:2008gr}, i.e.,
\begin{eqnarray}\label{eq:compactlag}
{\cal L}&=&ig\left\langle H^{(Q)}_b{\cal A}\!\!\!\slash_{ba}\gamma_5\overline{H}^{\,({Q})}_a\right\rangle+ig\left\langle \overline{H}^{(\overline{Q})}_a{\cal A}\!\!\!\slash_{ab}\gamma_5 H^{\,(\overline{Q})}_b\right\rangle\nonumber\\
&&+ik\left\langle T^{\,(Q)\mu}_b{\cal A}\!\!\!\slash_{ba}\gamma_5\overline{T}^{(Q)}_{a\mu}\right\rangle+ik\left\langle\overline{T}^{\,(\overline{Q})\mu}_a{\cal A}\!\!\!\slash_{ab}\gamma_5T^{(\overline{Q})}_{b\mu}\right\rangle\nonumber\\
&&+\left[i\left\langle T^{(Q)\mu}_b\left(\frac{h_1}{\Lambda_{\chi}}D_{\mu}{\cal A}\!\!\!\slash+\frac{h_2}{\Lambda_{\chi}}D\!\!\!\!/ {\cal A}_{\mu}\right)_{ba}\gamma_5\overline{H}^{\,(Q)}_a\right\rangle+h.c.\right]\nonumber\\
&&+\left[i\left\langle\overline{H}^{\,(\overline{Q})}_a\left(\frac{h_1}{\Lambda_{\chi}}{\cal A}\!\!\!\slash\stackrel{\leftarrow}{D_{\mu}'}+\frac{h_2}{\Lambda_{\chi}}{\cal A}_{\mu}\stackrel{\leftarrow}{D\!\!\!\slash'}\right)_{ab}\gamma_5T^{(\overline{Q})\mu}_b\right\rangle+h.c.\right]\nonumber\\
&&+\left\langle iH^{(Q)}_b\left(\beta v^{\mu}({\cal V}_{\mu}-\rho_{\mu})+\lambda \sigma^{\mu\nu}F_{\mu\nu}(\rho)\right)_{ba}\overline{H}^{\,(Q)}_a\right\rangle\nonumber\\
&&-\left\langle i\overline{H}^{(\overline{Q})}_a\left(\beta v^{\mu}({\cal V}_{\mu}-\rho_{\mu})-\lambda \sigma^{\mu\nu}F_{\mu\nu}(\rho)\right)_{ab}H^{\,(\overline{Q})}_b\right\rangle\nonumber\\
&&+\left\langle iT^{\,(Q)}_{b\lambda}\left(\beta^{\prime\prime} v^{\mu}({\cal V}_{\mu}-\rho_{\mu})+\lambda^{\prime\prime}\sigma^{\mu\nu}F_{\mu\nu}(\rho)\right)_{ba}\overline{T}^{(Q)\lambda}_{a}\right\rangle\nonumber\\
&&-\left\langle i\overline{T}^{\,(\overline{Q})}_{a\lambda}\left(\beta^{\prime\prime} v^{\mu}({\cal V}_{\mu}-\rho_{\mu})-\lambda^{\prime\prime}\sigma^{\mu\nu}F_{\mu\nu}(\rho)\right)_{ab}T^{(\overline{Q})\lambda}_{b}\right\rangle\nonumber\\
&&+\left[\left\langle T^{(Q)\mu}_b\left(i\zeta_1({\cal V}_{\mu}-\rho_{\mu})+\mu_{1}\gamma^{\nu}F_{\mu\nu}(\rho)\right)_{ba}\overline{H}^{\,(Q)}_a\right\rangle+h.c.\right]\nonumber\\
&&-\left[\left\langle\overline{H}^{\,(\overline{Q})}_a\left(i\zeta_1({\cal V}_{\mu}-\rho_{\mu})-\mu_1\gamma^{\nu}F_{\mu\nu}(\rho)\right)_{ab}T^{(\overline{Q})\mu}_b\right\rangle+h.c.\right],\nonumber
\end{eqnarray}
where the axial current ${\cal A}_{\mu}$, the vector current ${\cal V}_{\mu}$, and the vector meson field strength tensor $F_{\mu\nu}(\rho)$ are given by
\begin{eqnarray}
{\cal A}_{\mu}&=&\frac{1}{2}\left(\xi^{\dagger}\partial_{\mu}\xi-\xi\partial_{\mu}\xi^{\dagger}\right)_{\mu},\\
{\cal V}_{\mu}&=&\frac{1}{2}\left(\xi^{\dagger}\partial_{\mu}\xi+\xi\partial_{\mu}\xi^{\dagger}\right)_{\mu},\\
F_{\mu\nu}(\rho)&=&\partial_{\mu}\rho_{\nu}-\partial_{\nu}\rho_{\mu}+[\rho_{\mu},\rho_{\nu}],
\end{eqnarray}
respectively. Here, the pseudo-Goldstone and vector meson fields can be written as $\xi=\exp(i{\mathbb{P}}/f_{\pi})$ and $\rho_{\mu}=ig_V\mathbb{V}_{\mu}/\sqrt{2}$, respectively. The light pseudo-scalar meson matrix ${\mathbb{P}}$ and the light vector meson matrix $\mathbb{V}_{\mu}$ have the standard form, i.e.,
\begin{eqnarray}
\left.\begin{array}{c}
{\mathbb{P}} = {\left(\begin{array}{ccc}
       \frac{\pi^0}{\sqrt{2}}+\frac{\eta}{\sqrt{6}} &\pi^+ &K^+\\
       \pi^-       &-\frac{\pi^0}{\sqrt{2}}+\frac{\eta}{\sqrt{6}} &K^0\\
       K^-         &\bar K^0   &-\sqrt{\frac{2}{3}} \eta     \end{array}\right)},\\
{\mathbb{V}}_{\mu} = {\left(\begin{array}{ccc}
       \frac{\rho^0}{\sqrt{2}}+\frac{\omega}{\sqrt{2}} &\rho^+ &K^{*+}\\
       \rho^-       &-\frac{\rho^0}{\sqrt{2}}+\frac{\omega}{\sqrt{2}} &K^{*0}\\
       K^{*-}         &\bar K^{*0}   & \phi     \end{array}\right)}_{\mu}.
\end{array}\right.
\end{eqnarray}
In addition, the covariant derivatives  can be written as $D_{\mu}=\partial_{\mu}+{\cal V}_{\mu}$ and $D'_{\mu}=\partial_{\mu}-{\cal V}_{\mu}$.

With the above preparation, we can expand the compact effective Lagrangians to the leading order of the pseudo-Goldstone field. The expanded effective Lagrangians for the (anti-)charmed mesons and the exchanged light mesons are collected in Appendix~\ref{app00}. Here, the normalized relations for the pseudo-scalar charmed-strange meson $D_s$, the vector charmed-strange meson $D_s^{*}$, the axial-vector charmed-strange meson $D_{s1}$, and the tensor charmed-strange meson $D_{s2}^{*}$ can be expressed as
\begin{eqnarray}
\left.\begin{array}{ll}
\langle 0|D_s|c\bar{s}(0^-)\rangle=\sqrt{m_{D_s}},\quad&\langle 0|D_s^{*\mu}|c\bar{s}(1^-)\rangle=\epsilon^\mu\sqrt{m_{D_s^*}},\\
\langle 0|D_{s1}^{\mu}|c\bar{s}(1^+)\rangle=\epsilon^\mu\sqrt{m_{D_{s1}}},\quad&\langle 0|D_{s2}^{*\mu\nu}|c\bar{s}(2^+)\rangle=\zeta^{\mu\nu}\sqrt{m_{D_{s2}^*}},\\
\end{array}\right.\nonumber
\end{eqnarray}
respectively. Here, the polarization vector $\epsilon_{m}^{\mu}\,(m=0,\,\pm1)$ with spin-1 field is written as $\epsilon_{\pm}^{\mu}= \left(0,\,\pm1,\,i,\,0\right)/\sqrt{2}$ and $\epsilon_{0}^{\mu}= \left(0,0,0,-1\right)$ in the static limit, and the polarization tensor $\zeta_{m}^{\mu \nu}\,(m=0,\,\pm1,\,\pm2)$ with spin-2 field is constructed as $\zeta_{m}^{\mu \nu}=\sum_{m_1,m_2}\langle1,m_1;1,m_2|2,m\rangle\epsilon_{m_1}^{\mu}\epsilon_{m_2}^{\nu}$ \cite{Cheng:2010yd}.

\subsection{Effective potentials}\label{seb22}

\begin{figure}[!htbp]
\centering
\begin{tabular}{cc}
\includegraphics[width=0.17\textwidth]{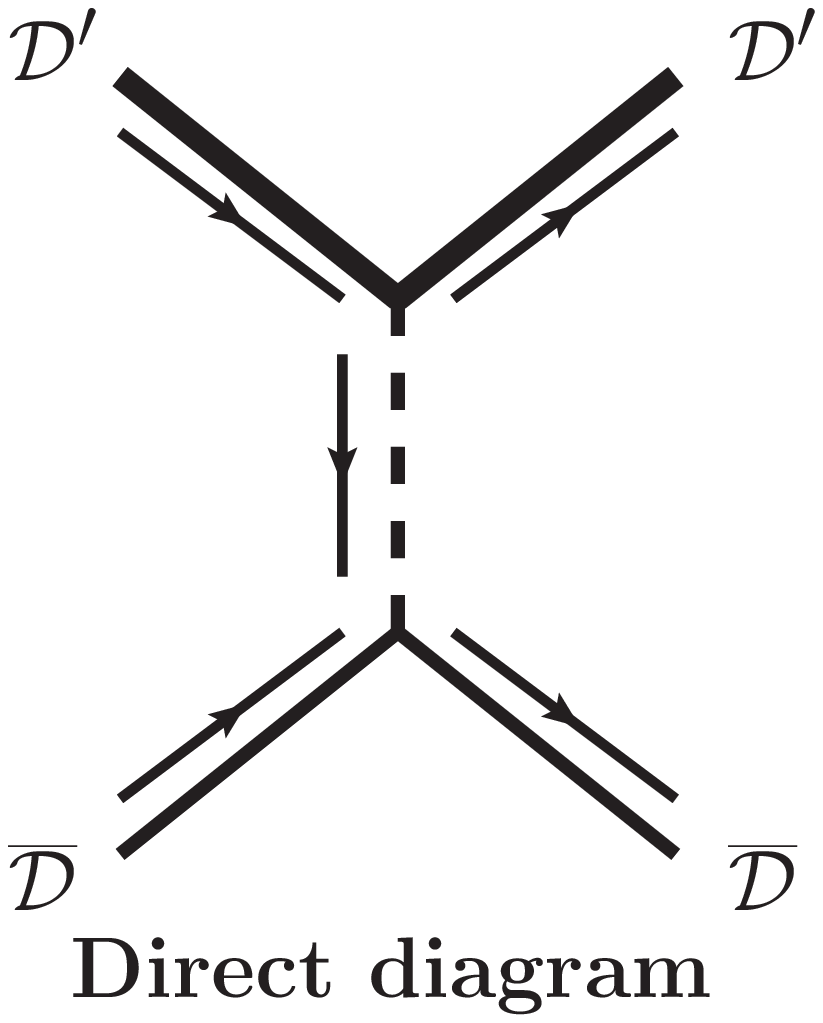}\quad\quad
\includegraphics[width=0.17\textwidth]{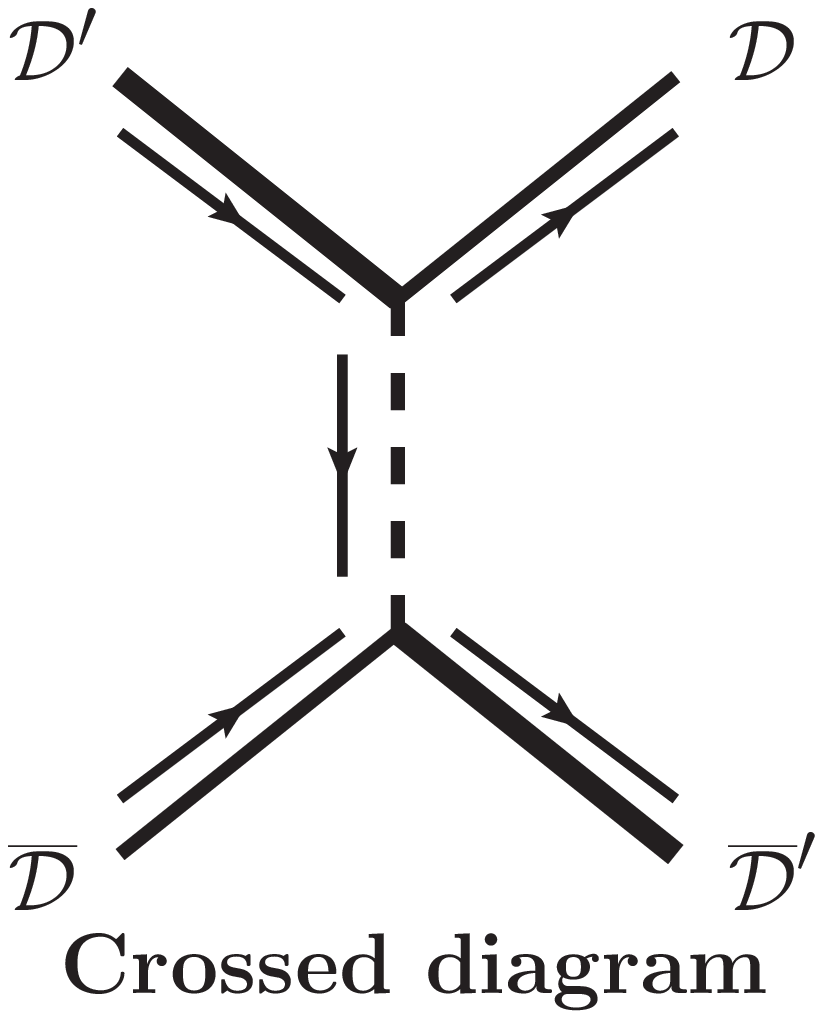}
\end{tabular}
\caption{The direct channel and crossed channel Feynman diagrams for the ${\cal D}^{\prime}\overline{{\cal D}}$ systems. The thick (thin) lines stand for the charmed-strange mesons ${\cal D}^{\prime}$(${\cal D}$), while the dashed lines represent the exchanged light mesons.}\label{fy}
\end{figure}

For getting the effective potentials in the coordinate space, we follow the standard strategy in Ref. \cite{Wang:2019nwt}.
Firstly, we write out the scattering amplitudes $\mathcal{M}(h_1h_2\to h_3h_4)$ of the involved scattering processes $h_1h_2\to h_3h_4$.
For the ${\cal D}^{\prime}\overline{{\cal D}}$ systems, there exist the direct channel and crossed channel  \cite{Liu:2007bf}, which are depicted in Fig.~\ref{fy}, where the notations $\cal{D}^{\prime}$ and $\cal{D}$ stand for two different charmed-strange mesons. {In Fig.~\ref{fy}, according to requirement of the spin-parity conservation, we may determine the exchanged particles for certain hadron-hadron system in the framework of the OBE model. Here, we take the $D_s\bar D_{s1}$ system as an example to illustrate it. Since the $D_sD_s\eta$ and $D_sD_{s1}\eta$ vertexes are strictly forbidden by the spin-parity conservation, there only exists the $\phi$ exchange contribution to the direct and crossed channels for the $D_s\bar D_{s1}$ system.}

Secondly, we have
\begin{eqnarray}
\mathcal{V}^{h_1h_2\to h_3h_4}_E(\bm{q})=-\frac{\mathcal{M}(h_1h_2\to h_3h_4)}{\sqrt{\prod_i 2 m_i}}
\end{eqnarray}
with the Breit approximation \cite{Breit:1929zz,Breit:1930zza} and the non-relativistic normalization. Here, $\mathcal{V}^{h_1h_2\to h_3h_4}_E(\bm{q})$ is the effective potentials in the momentum space, and $m_{i}\,(i=h_1,\,h_2,\,h_3,\,h_4)$ represent the masses of the initial and final states.

Thirdly, we need to get the effective potentials in the coordinate space $\mathcal{V}^{h_1h_2\to h_3h_4}_E(\bm{r})$ by performing the Fourier transformation to $\mathcal{V}^{h_1h_2\to h_3h_4}_E(\bm{q})$, i.e.,
\begin{eqnarray}
\mathcal{V}^{h_1h_2\to h_3h_4}_E(\bm{r}) =\int \frac{d^3\bm{q}}{(2\pi)^3}e^{i\bm{q}\cdot\bm{r}}\mathcal{V}^{h_1h_2\to h_3h_4}_E(\bm{q})\mathcal{F}^2(q^2,m_E^2).\nonumber
\end{eqnarray}
For compensating the off-shell effects of the exchanged light mesons and reflecting the inner structures of the interaction vertex, the form factor should be introduced in every interaction vertex \cite{Tornqvist:1993ng,Tornqvist:1993vu}. We should indicate that the form factor also plays the role to regulate the effective potentials in the coordinate space since these effective potentials in the coordinate space have the singular delta-function terms \cite{Wang:2019nwt}.
In this work, we introduce the monopole type form factor $\mathcal{F}(q^2,m_E^2) = (\Lambda^2-m_E^2)/(\Lambda^2-q^2)$ in the OBE model \cite{Tornqvist:1993ng,Tornqvist:1993vu}. Here, $\Lambda$, $m_E$, and $q$ are the cutoff parameter, the mass, and the four momentum of the exchanged light mesons, respectively.

In order to obtain the effective potentials of these focused systems, we need to construct the flavor and spin-orbital wave functions of systems. For the hidden-charm and hidden-strange ${\cal D}^{\prime}\overline{{\cal D}}$ molecular systems, we need distinguish the charge parity quantum numbers due to the charge conjugate transformation invariance. The flavor wave function $|I,I_{3}\rangle$ can be expressed as $\left|0,0\right\rangle =\left|{\cal D}^{\prime+} {\cal D}^{-}+c{\cal D}^{+} {\cal D}^{\prime-}\right\rangle/\sqrt{2}$ \cite{Liu:2008fh,Liu:2008tn,Sun:2012sy}. {Here, we want to emphasize that there exists a relation of $c$ and the $C$ parity, i.e., $C=-c\cdot(-1)^{J_1-J_2-J}$,} where the notations $J$, $J_1$, and $J_2$ stand for the total angular momentum quantum numbers of the ${\cal D}^{\prime}\overline{{\cal D}}$, the charmed-strange mesons $\cal{D}^{\prime}$, and the charmed-strange mesons $\cal{D}$, respectively.

And then, we may get the spin-orbital wave functions $|{}^{2S+1}L_{J}\rangle$ for all of the investigated hidden-charm and hidden-strange molecular systems, i.e.,
\begin{eqnarray}
|{\cal D}_{0}{\overline {\cal D}}_{1}\rangle &=&\sum_{m,m_L}C^{J,M}_{1m,Lm_L}\epsilon_{m}^\mu|Y_{L,m_L}\rangle,\\
|{\cal D}_{0}{\overline {\cal D}}_{2}\rangle &=&\sum_{m,m_L}C^{J,M}_{2m,Lm_L}\zeta_{m}^{\mu\nu}|Y_{L,m_L}\rangle,\\
|{\cal D}_{1}{\overline {\cal D}}_{1}\rangle &=&\sum_{m,m^{\prime},m_S,m_L}C^{S,m_S}_{1m,1m^{\prime}}C^{J,M}_{Sm_S,Lm_L}\epsilon_{m}^\mu\epsilon_{m^{\prime}}^\nu|Y_{L,m_L}\rangle,\\
|{\cal D}_{1}{\overline {\cal D}}_{2}\rangle &=& \sum_{m,m^{\prime},m_S,m_L}C^{S,m_S}_{1m,2m^{\prime}}C^{J,M}_{Sm_S,Lm_L}\epsilon_{m}^\lambda\zeta_{m^{\prime}}^{\mu\nu}|Y_{L,m_L}\rangle.
\end{eqnarray}
In the above expressions, the notations ${\cal D}_{0}$, ${\cal D}_{1}$, and ${\cal D}_{2}$ denote the charmed-strange mesons with the total angular momentum quantum numbers $J=0$, $1$, and $2$, respectively. The constant $C^{e,f}_{ab,cd}$ is the Clebsch-Gordan coefficient, and $|Y_{L,m_L}\rangle$ stands for the spherical harmonics function.

Through the above preparation, we can derive the effective potentials in the coordinate space for all of the investigated hidden-charm and hidden-strange systems, which are shown in Appendix~\ref{app01}. We need to emphasize that the total effective potentials contain the direct channel and crossed channel contributions, which can be written in a general form, i.e.,
\begin{eqnarray}
\mathcal{V}_{Total}(\bm{r})=\mathcal{V}_{D}(\bm{r})+c\,\mathcal{V}_{C}(\bm{r}),
\end{eqnarray}
where $\mathcal{V}_{D}(\bm{r})$ and $\mathcal{V}_{C}(\bm{r})$ are the effective potentials corresponding to the direct channel and crossed channel, respectively.

In the present work, we will also discuss the bound state properties of the $S$-wave $H_s T_s$ systems.
Their OBE effective potentials can be related to the effective potentials of  the $H_s \bar T_s$ systems  by the $G$-parity rule \cite{Klempt:2002ap}, i.e.,
\begin{equation}
V^{H_s T_s \rightarrow H_s T_s}({\bf{r}})=\sum_{i}G_i V^{H_s \bar T_s \rightarrow H_s \bar T_s}_{i}({\bf{r}}),
\end{equation}
where $G_{i}$ is the $G$-parity of some exchanged mesons. We need to emphasize that we only need to consider the direct diagram contribution to the $H_s T_s$ systems since the charge conjugate transformation invariance for the $H_s T_s$ systems does not exist.

\section{Numerical Results}\label{sec3}

For describing the interactions quantitatively, we need the values of the coupling constants. The pionic coupling constant $g$ can be determined by reproducing the experimental width of the process $D^{*+} \to D^{0}\pi^{+}$ \cite{Falk:1992cx,Isola:2003fh}. With the available experimental information, the authors extracted the coupling constant $h^{\prime}=(h_1+h_2)/\Lambda_{\chi}$ \cite{Casalbuoni:1996pg}. According to the vector meson dominance mechanism \cite{Isola:2003fh}, the coupling constants $\beta$, $\zeta_1$, $\mu_1$, and $g_V$ can be obtained \cite{Isola:2003fh,Cleven:2016qbn}. Among them, the coupling constants $\zeta_1$ and $\mu_1$ are consistent with the numerical results in Refs. \cite{Dong:2019ofp,He:2019csk}. The coupling constant $\lambda$ may be fixed by comparing the form factor obtained from the lattice QCD with this calculated via the light cone sum rule \cite{Isola:2003fh}. In addition, the coupling constants with the charmed meson in $T$-doublet can be estimated with the quark model in Refs. \cite{Wang:2019aoc,Wang:2019nwt,Wang:2020lua}. Meanwhile, the corresponding phase factors between these coupling constants related to the effective potentials in the direct channels are fixed with the quark model \cite{Riska:2000gd}. In addition, we also need the parameters of the hadron masses \cite{Tanabashi:2018oca}. The values of the coupling constants and the hadron masses are listed in Table~\ref{Parameters}. Thus, the cutoff $\Lambda$ in the form factor is only one free parameter in our numerical analysis, we attempt to find the loosely bound solutions by varying the cutoff parameters $\Lambda$ from 1 to 5$~{\rm GeV}$ in the following\footnote{{For the $Y(4274)$ \cite{Aaij:2016nsc}, the spin-parity quantum number $J^{PC}=1^{++}$ was measured by the LHCb Collaboration later \cite{Aaij:2016iza}, which is not consistent with the $J^{PC}$ quantum number corresponding to the assignment of the $S$-wave $D_s\bar{D}_{s0}(2317)$ molecular state even. Considering this situation, we did not adopt the observed $Y(4274)$ to fix the Lambda cutoff in this work. As is well known, the deuteron is a loosely bound state composed of a proton and a neutron, and may be regarded as an ideal molecular state. By reproducing the binding energy and root-mean-square radius of  deuteron with the OBE potential model, the $\Lambda$ cutoff should be around 1 GeV \cite{Tornqvist:1993ng,Tornqvist:1993vu,Wang:2019nwt}}}. In this work, we take cutoff range around 1 GeV to discuss possible hadron molecular states with hidden-charm and hidden-strange.
\renewcommand\tabcolsep{0.10cm}
\renewcommand{\arraystretch}{1.50}
\begin{table}[!htbp]
\caption{The summary of the coupling constants and the hadron masses adopted in our calculations. Unit of the coupling constants $h^{\prime}$, $\lambda$, $\lambda^{\prime\prime}$, and $\mu_1$ is $\rm{GeV}^{-1}$, and the coupling constant $f_\pi$ is given in units of $\rm{GeV}$.}\label{Parameters}
\begin{tabular}{c|cccc}\toprule[1pt]\toprule[1pt]
\multirow{4}{*}{Coupling constants}&$(g,\,k)$&$|h^{\prime}|$ &$f_\pi$&$(\beta,\,\beta^{\prime\prime})$\\
                                    &$(0.59,\,0.59)$&0.55&0.132&$(-0.90,0.90)$\\
                                    &$(\lambda,\,\lambda^{\prime\prime})$ &$|\zeta_1|$&$\mu_1$ &$g_V$\\
                                    &$(-0.56,\,0.56)$&0.20&0&5.83\\\midrule[1.0pt]
\multirow{4}{*}{Hadron masses}&$\eta$&$\phi$&$D_s$&$D_s^*$\\
\multirow{3}{*}{($\rm{MeV}$)}&$547.86$&1019.46&1968.34&$2112.20$\\
                                    &$D_{s1}$&$D_{s2}^*$&   &  \\
                                    &$2535.11$&2569.10&  & \\
\bottomrule[1pt]\bottomrule[1pt]
\end{tabular}
\end{table}

\subsection{The hidden-charm and hidden-strange molecular systems}\label{seb01}

The hadronic molecular state is a loosely bound state, where the binding energy should be tens of MeV, and the typical size should be larger than the size of all the included hadrons \cite{Chen:2017xat}. The above criteria may provide us the critical information to identify the hadronic molecular candidates. Besides, it is important to note that the $S$-wave bound states should firstly appear since there exist the repulsive centrifugal potential $\ell(\ell+1)/2\mu r^2$ for the higher partial wave states ($\mu$ and $\ell$ respectively are the reduced mass and the angular momentum quantum number for the investigated system). Thus, we are mainly interested in the $S$-wave $H_s\bar T_s$ systems in this work.

In fact, the $S$-$D$ wave mixing effect may play an important role to modify the bound properties of the deuteron,  which may be regarded as an ideal molecular state \cite{Tornqvist:1993ng,Tornqvist:1993vu,Wang:2019nwt}. In this work, we also consider the $S$-$D$ wave mixing effect to the $S$-wave $H_s\bar T_s$ systems. The relevant spin-orbit wave functions  $|{}^{2S+1}L_{J}\rangle$ are summarized in Table~\ref{spin-orbit wave functions}, where $S$, $L$, and $J$ denote the spin, angular momentum, and total angular momentum quantum numbers, respectively.
\renewcommand\tabcolsep{0.18cm}
\renewcommand{\arraystretch}{1.50}
\begin{table}[!htpb]
\centering
\caption{The relevant quantum numbers $J^{P}$ and possible channels $|{}^{2S+1}L_{J}\rangle$ involved in the $S$-wave $H_s\bar T_s$ systems. Here, ``$...$" means that the $S$-wave components for the corresponding channels do not exist.}\label{spin-orbit wave functions}
\begin{tabular}{c|cccc}\toprule[1pt]\toprule[1pt]
 $J^{P}$&$D_s\bar D_{s1}$&$D_s\bar D_{s2}^{*}$&$D_s^{*}\bar D_{s1}$&$D_s^{*}\bar D_{s2}^{*}$\\\midrule[1.0pt]
$0^{-}$&$...$&$...$&$|{}^1\mathbb{S}_{0}\rangle/|{}^5\mathbb{D}_{0}\rangle$&$...$\\
$1^{-}$ &$|{}^3\mathbb{S}_{1}\rangle/|{}^3\mathbb{D}_{1}\rangle$&$...$&$|{}^3\mathbb{S}_{1}\rangle/|{}^{3,5}\mathbb{D}_{1}\rangle$&$|{}^3\mathbb{S}_{1}\rangle/|{}^{3,5,7}\mathbb{D}_{1}\rangle$\\
$2^{-}$&$...$&$|{}^5\mathbb{S}_{2}\rangle/|{}^5\mathbb{D}_{2}\rangle$&$|{}^5\mathbb{S}_{2}\rangle/|{}^{1,3,5}\mathbb{D}_{2}\rangle$&$|{}^5\mathbb{S}_{2}\rangle/|{}^{3,5,7}\mathbb{D}_{2}\rangle$\\
$3^{-}$&$...$&$...$&$...$&$|{}^7\mathbb{S}_{3}\rangle/|{}^{3,5,7}\mathbb{D}_{3}\rangle$\\
\bottomrule[1pt]\bottomrule[1pt]
\end{tabular}
\end{table}

By solving the Schr$\ddot{\rm o}$dinger equation
\begin{eqnarray}
-\frac{1}{2\mu}\left(\nabla^2-\frac{
\ell(\ell+1)}{r^2}\right)\psi(r)+V(r)\psi(r)=E\psi(r),
\end{eqnarray}
we can find bound state solutions of these discussed systems.
Here, $\nabla^2=\frac{1}{r^2}\frac{\partial}{\partial r}r^2\frac{\partial}{\partial r}$ and $\mu=\frac{m_1m_2}{m_1+m_2}$ as the reduced mass for the investigated system. The bound state solutions include the binding energy $E$ and the radial wave function $\psi(r)$. In addition, we can further calculate the root-mean-square radius $r_{\rm RMS}$ and the probability of the individual channel $P_i$. In the following, we present the numerical results for single channel and coupled-channel cases, respectively.

\subsubsection{The single channel analysis}

In our numerical analysis, we firstly give the results without considering the $S$-$D$ wave mixing effect. After that, we further take into account the $S$-$D$ wave mixing effect, and repeat the numerical analysis. For the $S$-wave $H_s \bar T_s$ systems, the relevant numerical results are collected in Tables~\ref{sr1}-\ref{sr4} within the OBE model, which include the cutoff parameter $\Lambda$, the binding energy $E$, the root-mean-square radius $r_{\rm RMS}$, and the probability of the individual channel $P_i$. 

Since the $D_sD_s\eta$ and $D_sD_{s1}\eta$ vertexes are forbidden by the spin-parity conservation, there only exist the $\phi$ exchange contribution to the direct and crossed channels for the $D_s\bar D_{s1}$ system. By performing numerical calculations, we can find that there exist the loosely bound state solutions for the $S$-wave $D_s\bar D_{s1}$ states with $J^{PC}=1^{--}$ and $1^{-+}$ when the cutoff parameters $\Lambda$ are larger than 4.56 GeV and 3.85 GeV, respectively. However, such cutoff parameters are unusual and deviate from the reasonable range around 1.00 GeV \cite{Tornqvist:1993ng,Tornqvist:1993vu,Wang:2019nwt}, which reflects that the $\phi$ exchange interaction is not strong enough to generate the bound states for the $S$-wave $D_s\bar D_{s1}$ states with $J^{PC}=1^{--}$ and $1^{-+}$. Thus, these states as the candidates of the hadronic molecular states are no priority. Besides, we also notice that the $D_s\bar D_{s1}$ system without and with the $S$-$D$ wave mixing effect have the same bound state properties in our calculation, which is not surprising since the contribution of the tensor forces from the $S$-$D$ wave mixing effect for the $D_s\bar D_{s1}$ interactions disappears.
\renewcommand\tabcolsep{0.17cm}
\renewcommand{\arraystretch}{1.50}
\begin{table}[!htbp]
\caption{Bound state solutions for the $S$-wave $D_s\bar D_{s1}$  system. The cutoff $\Lambda$, the binding energy $E$, and the root-mean-square radius $r_{RMS}$ are in units of $ \rm{GeV}$, $\rm {MeV}$, and $\rm {fm}$, respectively. Here, we label the major probability for the corresponding channels in a bold manner, and the second column shows the numerical results without considering the $S$-$D$ wave mixing effect while the last column shows the relevant results with the $S$-$D$ wave mixing effect.}\label{sr1}
\begin{tabular}{c|ccc|cccc}\toprule[1pt]\toprule[1pt]
$J^{PC}$&$\Lambda$ &$E$&$r_{\rm RMS}$&$\Lambda$ &$E$&$r_{\rm RMS}$&$P({}^3\mathbb{S}_{1}/{}^3\mathbb{D}_{1})$ \\
\cline{1-8}
\multirow{3}{*}{$1^{--}$} &4.56&$-0.30$&4.73       &4.56&$-0.30$&4.73&\textbf{100.00}/$o(0)$\\
                             &4.78&$-0.67$&3.63        &4.78&$-0.67$&3.63&\textbf{100.00}/$o(0)$\\
                             &5.00&$-1.15$&2.89        &5.00&$-1.15$&2.89&\textbf{100.00}/$o(0)$\\
\cline{1-8}
\multirow{3}{*}{$1^{-+}$}  &3.85&$-0.31$&4.72         &3.85&$-0.31$&4.72&\textbf{100.00}/$o(0)$\\
                             &4.43&$-2.26$&2.12    &4.43&$-2.26$&2.12&\textbf{100.00}/$o(0)$\\
                             &5.00&$-5.07$&1.46    &5.00&$-5.07$&1.46&\textbf{100.00}/$o(0)$\\
\bottomrule[1pt]\bottomrule[1pt]
\end{tabular}
\end{table}

For the $S$-wave $D_s\bar D_{s2}^{*}$ state with $J^{PC}=2^{--}$, there do not exist the loosely bound state solutions if we only consider the $S$-wave component until we increase the cutoff parameter $\Lambda$ to be around 5$~{\rm GeV}$. There exist weakly bound state solutions when the value of the cutoff parameter is taken around 4.70 GeV if adding the contributions of the $S$-$D$ wave mixing effect. For the $S$-wave $D_s\bar D_{s2}^{*}$ state with $J^{PC}=2^{-+}$, we can find the loosely bound state solutions when the cutoff parameter $\Lambda$ larger than 2.40 GeV, even if the $S$-$D$ wave mixing effect is included in our calculation. According to our quantitative analysis, it is obvious that the corresponding cutoff parameters are far away from the usual value around 1 GeV for the $S$-wave $D_s\bar D_{s2}^{*}$ bound states with $J^{PC}=2^{--}$ and $2^{-+}$ \cite{Tornqvist:1993ng,Tornqvist:1993vu,Wang:2019nwt}. Here, the large cutoff parameter means that the attractive forces are not strong enough to form these loosely bound states. Thus, we conclude that our numerical results disfavor the existence of the hadronic molecular state candidates for the $S$-wave $D_s\bar D_{s2}^{*}$ states with $J^{PC}=2^{--}$ and $2^{-+}$.
\renewcommand\tabcolsep{0.15cm}
\renewcommand{\arraystretch}{1.50}
\begin{table}[!htbp]
\caption{Bound state solutions for the $S$-wave $D_s\bar D_{s2}^{*}$ system. Conventions are the same as Table~\ref{sr1}. Here, ``$\times$" indicates no binding energy when scanning the $\Lambda$ range 1-5 ${\rm GeV}$.}\label{sr2}
\begin{tabular}{c|ccc|cccc}\toprule[1pt]\toprule[1pt]
$J^{PC}$&$\Lambda$ &$E$&$r_{\rm RMS}$&$\Lambda$ &$E$&$r_{\rm RMS}$&$P({}^5\mathbb{S}_{2}/{}^5\mathbb{D}_{2})$ \\
\cline{1-8}
\multirow{3}{*}{$2^{--}$}    &$\times$&$\times$&$\times$     &4.70&$-0.22$&5.00&\textbf{99.87}/0.13\\
                             &$\times$&$\times$&$\times$     &4.76&$-2.87$&1.81&\textbf{99.53}/0.47\\
                             &$\times$&$\times$&$\times$     &4.82&$-9.76$&1.00&\textbf{99.07}/0.93\\
\cline{1-8}
\multirow{3}{*}{$2^{-+}$}     &2.44&$-0.31$&4.69      &2.41&$-0.34$&4.58&\textbf{99.99}/0.01\\
                             &2.52&$-2.95$&1.87       &2.49&$-3.35$&1.76&\textbf{99.97}/0.03\\
                             &2.60&$-10.16$&1.06      &2.56&$-10.26$&1.05&\textbf{99.93}/0.07\\
\bottomrule[1pt]\bottomrule[1pt]
\end{tabular}
\end{table}

In this work, we need to focus on the ${\bm{q}}^4$ correction terms in the effective potentials, the expressions of these correction terms are a little tricky (see Eq. (\ref{correction terms}) of the Appendix~\ref{app01} for more details). Through our numerical analysis for the $D_s\bar D_{s2}^{*}$ system, we can obtain an inequality $\Lambda\left(J^{PC}=2^{--}\right)>\Lambda\left(J^{PC}=2^{-+}\right)$ when taking the same binding energy. This difference is caused by the ${\bm{q}}^4$ correction terms in the effective potentials. Usually, a loosely bound state with smaller cutoff parameter corresponds to the more attractive interaction, which means that the ${\bm{q}}^4$ correction terms in the effective potentials are favorable for forming the $D_s\bar D_{s2}^{*}$ molecular state with $J^{PC}=2^{-+}$. In contrast, these correction terms are unfavorable for forming the $D_s\bar D_{s2}^{*}$ state with $J^{PC}=2^{--}$. Based on the analysis mentioned above, it is clear that the ${\bm{q}}^4$ correction terms in the interactions play an important role to modify the behavior of the loosely bound state in some cases.

Besides the $S$-wave $D_s\bar D_{s1}(\bar D_{s2}^{*})$ systems, we also investigate the bound state properties of the $S$-wave $D_s^{*}\bar D_{s1}(\bar D_{s2}^{*})$ systems in the current work. For the $D_s^{*}\bar D_{s1}(\bar D_{s2}^{*})$ systems, we notice that the $\eta$ and $\phi$ exchange contributions to the direct and crossed diagrams are also allowed, and their interactions are simultaneously associated with the total angular momentum $J$ and the charge parity $C$.
\renewcommand\tabcolsep{0.06cm}
\renewcommand{\arraystretch}{1.50}
\begin{table}[!htbp]
\caption{Bound state solutions for the $S$-wave $D_s^{*}\bar D_{s1}$  system. Conventions are the same as Table~\ref{sr1}.}\label{sr3}
\begin{tabular}{c|ccc|cccc}\toprule[1pt]\toprule[1pt]
$J^{PC}$&$\Lambda$ &$E$&$r_{\rm RMS}$&$\Lambda$ &$E$&$r_{\rm RMS}$&$P({}^1\mathbb{S}_{0}/{}^5\mathbb{D}_{0})$ \\
\cline{1-8}
\multirow{3}{*}{$0^{--}$}        &1.68&$-0.41$&4.14      &1.68&$-0.42$&4.09&\textbf{100.00}/$o(0)$\\
                             &1.72&$-4.59$&1.38      &1.72&$-4.63$&1.38&\textbf{100.00}/$o(0)$\\
                             &1.75&$-10.27$&0.95      &1.75&$-10.32$&0.95&\textbf{100.00}/$o(0)$\\
\cline{1-8}
\multirow{3}{*}{$0^{-+}$}        &1.55&$-0.22$&4.98      &1.55&$-0.35$&4.48&\textbf{99.95}/0.05\\
                             &1.59&$-3.91$&1.57      &1.59&$-4.28$&1.51&\textbf{99.89}/0.11\\
                             &1.62&$-9.75$&1.03      &1.62&$-10.26$&1.01&\textbf{99.87}/0.13\\\midrule[1.0pt]
$J^{PC}$&$\Lambda$ &$E$&$r_{\rm RMS}$&$\Lambda$ &$E$&$r_{\rm RMS}$&$P({}^3\mathbb{S}_{1}/{}^3\mathbb{D}_{1}/{}^5\mathbb{D}_{1})$ \\
\multirow{3}{*}{$1^{-+}$}        &1.83&$-0.30$&4.63     &1.82&$-0.23$&4.95&\textbf{99.96}/0.04/$o(0)$\\
                             &1.89&$-3.84$&1.59     &1.88&$-3.54$&1.66&\textbf{99.89}/0.11/$o(0)$\\
                             &1.94&$-10.06$&1.03     &1.93&$-9.53$&1.05&\textbf{99.86}/0.13/0.01\\
\cline{1-8}
\multirow{3}{*}{$1^{--}$}         &2.00&$-0.32$&4.51     &1.99&$-0.23$&4.90&\textbf{100.00}/$o(0)$/$o(0)$\\
                             &2.07&$-4.20$&1.47      &2.06&$-4.01$&1.51&\textbf{99.99}/$o(0)$/0.01\\
                             &2.13&$-10.82$&0.95     &2.12&$-10.71$&0.95&\textbf{99.97}/0.01/0.02\\\midrule[1.0pt]
$J^{PC}$&$\Lambda$ &$E$&$r_{\rm RMS}$&$\Lambda$ &$E$&$r_{\rm RMS}$&$P({}^5\mathbb{S}_{2}/{}^1\mathbb{D}_{2}/{}^3\mathbb{D}_{2}/{}^5\mathbb{D}_{2})$ \\
\multirow{3}{*}{$2^{--}$}         &3.313&$-0.33$&4.46      &3.119&$-0.29$&4.67&\textbf{99.55}/0.40/$o(0)$/0.05\\
                             &3.316&$-5.60$&1.23       &3.125&$-3.81$&1.55&\textbf{98.28}/1.59/$o(0)$/0.13\\
                             &3.318&$-13.00$&0.78       &3.130&$-10.75$&0.92&\textbf{97.03}/2.81/$o(0)$/0.16\\
\cline{1-8}
\multirow{3}{*}{$2^{-+}$}         &2.96&$-0.29$&4.84      &2.86&$-0.31$&4.76&\textbf{99.97}/0.02/$o(0)$/0.01\\
                             &3.20&$-3.27$&1.87       &3.04&$-3.07$&1.92&\textbf{99.85}/0.09/$o(0)$/0.06\\
                             &3.43&$-10.00$&1.17       &3.22&$-10.20$&1.15&\textbf{99.59}/0.21/$o(0)$/0.20\\
\bottomrule[1pt]\bottomrule[1pt]
\end{tabular}
\end{table}
\renewcommand\tabcolsep{0.07cm}
\renewcommand{\arraystretch}{1.50}
\begin{table}[!htbp]
\caption{Bound state solutions for the $S$-wave $D_s^{*}\bar D_{s2}^{*}$ system. Conventions are the same as Table~\ref{sr1}.}\label{sr4}
\begin{tabular}{c|ccc|cccc}\toprule[1pt]\toprule[1pt]
$J^{PC}$&$\Lambda$ &$E$&$r_{\rm RMS}$&$\Lambda$ &$E$&$r_{\rm RMS}$&$P({}^3\mathbb{S}_{1}/{}^3\mathbb{D}_{1}/{}^5\mathbb{D}_{1}/{}^7\mathbb{D}_{1})$ \\
\cline{1-8}
\multirow{3}{*}{$1^{--}$}        &1.70&$-0.60$&3.65     &1.69&$-0.50$&3.90&\textbf{99.97}/0.02/$o(0)$/0.01\\
                             &1.74&$-4.15$&1.49     &1.73&$-4.01$&1.51&\textbf{99.94}/0.05/$o(0)$/0.01\\
                             &1.78&$-10.68$&0.96     &1.77&$-10.69$&0.96&\textbf{99.91}/0.08/$o(0)$/0.01\\
\cline{1-8}
\multirow{3}{*}{$1^{-+}$}        &1.61&$-0.56$&3.76     &1.60&$-0.28$&4.67&\textbf{99.99}/$o(0)$/$o(0)$/0.01\\
                             &1.65&$-5.12$&1.36     &1.64&$-4.28$&1.48&\textbf{99.96}/0.01/$o(0)$/0.03\\
                             &1.68&$-11.56$&0.93     &1.67&$-10.37$&0.98&\textbf{99.94}/0.02/$o(0)$/0.04\\\midrule[1.0pt]
$J^{PC}$&$\Lambda$ &$E$&$r_{\rm RMS}$&$\Lambda$ &$E$&$r_{\rm RMS}$&$P({}^5\mathbb{S}_{2}/{}^3\mathbb{D}_{2}/{}^5\mathbb{D}_{2}/{}^7\mathbb{D}_{2})$ \\
\multirow{3}{*}{$2^{-+}$}         &2.67&$-0.28$&4.69     &2.63&$-0.26$&4.77&\textbf{99.99}/$o(0)$/0.01/$o(0)$\\
                             &2.86&$-3.58$&1.60     &2.80&$-3.68$&1.58&\textbf{99.96}/$o(0)$/0.04/$o(0)$\\
                             &3.05&$-10.08$&0.98     &2.96&$-10.43$&0.96&\textbf{99.91}/$o(0)$/0.09/$o(0)$\\
\cline{1-8}
\multirow{3}{*}{$2^{--}$}         &1.91&$-0.47$&4.06     &1.90&$-0.26$&4.82&\textbf{99.99}/$o(0)$/0.01/$o(0)$\\
                             &1.96&$-3.57$&1.65     &1.96&$-3.81$&1.60&\textbf{99.97}/$o(0)$/0.03/$o(0)$\\
                             &2.01&$-10.20$&1.01     &2.01&$-10.62$&1.00&\textbf{99.96}/$o(0)$/0.04/$o(0)$\\\midrule[1.0pt]
$J^{PC}$&$\Lambda$ &$E$&$r_{\rm RMS}$&$\Lambda$ &$E$&$r_{\rm RMS}$&$P({}^7\mathbb{S}_{3}/{}^3\mathbb{D}_{3}/{}^5\mathbb{D}_{3}/{}^7\mathbb{D}_{3})$ \\
\multirow{3}{*}{$3^{--}$}         &4.18&$-0.28$&4.87     &3.87&$-0.29$&4.80&\textbf{99.96}/0.03/$o(0)$/0.01\\
                              &4.59&$-2.56$&2.07     &4.04&$-1.97$&2.32&\textbf{99.83}/0.13/$o(0)$/0.04\\
                              &5.00&$-6.36$&1.40     &4.20&$-9.89$&1.15&\textbf{98.87}/0.97/$o(0)$/0.16\\
\bottomrule[1pt]\bottomrule[1pt]
\end{tabular}
\end{table}

From the numerical results for the $S$-wave $D_s^{*}\bar D_{s1}$ and $D_s^{*}\bar D_{s2}^{*}$ systems with and without considering the $S$-$D$ wave mixing effect, we can find several interesting results:
\begin{enumerate}
  \item When tuning the cutoff parameters $\Lambda$ from 1 GeV to 5 GeV, we can obtain the loosely bound solutions for the $D_s^{*}\bar D_{s1}$ states with $J^{PC}$=$0^{--}, 0^{-+}, 1^{--}, 1^{-+}, 2^{--}, 2^{-+}$ and the $D_s^{*}\bar D_{s2}^{*}$ states with $J^{PC}$=$1^{--}, 1^{-+}, 2^{--}, 2^{-+}, 3^{--}$, where the $S$-wave $D_s^{*}\bar D_{s1}$ and $D_s^{*}\bar D_{s2}^{*}$ states with lower spin can be bound more tightly;

  \item If strictly considering this criterion of the cutoff value $\Lambda$ around 1 GeV \cite{Tornqvist:1993ng,Tornqvist:1993vu,Wang:2019nwt}, our results disfavor the existence of the hidden-charm and hidden-strange tetraquark molecular candidates for the $S$-wave $D_s^{(*)}\bar D_{s1}(\bar D_{s2}^{*})$ states;

  \item If the cutoff $\Lambda$ smaller than 1.70 GeV is a reasonable input parameter\footnote{{In this work, we take a $\Lambda$ range from 1 GeV to 5 GeV when finding the bound state solution of these discussed systems. According to the experience of studying deuteron, the cutoff value should not be far away from 1 GeV. Thus, in this work, we take such so-called criteria to make discussion of possible molecular states.}}, our results suggest that the $D_s^{*}\bar D_{s1}$ states with $J^{PC}$=$0^{--}, 0^{-+}$ and the $D_s^{*}\bar D_{s2}^{*}$ states with $J^{PC}$=$1^{--}, 1^{-+}$ are good candidates of the hidden-charm and hidden-strange molecular states. Here, we need to emphasize that the $D_s^{*}\bar D_{s1}$ state with $J^{PC}$=$0^{--}$ and the $D_s^{*}\bar D_{s2}^{*}$ state with $J^{PC}$=$1^{-+}$  are very different from the $D_s^{*}\bar D_{s1}$ state with $J^{PC}$=$0^{-+}$ and the $D_s^{*}\bar D_{s2}^{*}$ state with $J^{PC}$=$1^{--}$, since they have the exotic spin-parity quantum numbers apparently different from the conventional hadrons.

  \item The $S$-$D$ wave mixing effect plays a minor role in generating the $S$-wave $D_s^{*}\bar D_{s1}$ and $D_s^{*}\bar D_{s2}^{*}$ bound states in many cases.
\end{enumerate}

\subsubsection{The coupled-channel analysis}

In this subsection, we discuss the $H_s \bar T_s$ systems with $J = 1, 2$ by considering the coupled-channel effect. In the coupled-channel approach, we need to emphasize that the binding energy $E$ is determined by the lowest mass threshold among various involved channels \cite{Chen:2017xat}. The relevant numerical results for the $H_s \bar T_s$ coupled systems with $J^{PC}=1^{--},\,1^{-+},\,2^{--},\,2^{-+}$ for the different lowest mass thresholds are given in Table \ref{cr1}.
\renewcommand\tabcolsep{0.09cm}
\renewcommand{\arraystretch}{1.50}
\begin{table}[!htbp]
\caption{Bound state solutions for the $S$-wave $H_s \bar T_s$ coupled systems. Conventions are the same as Table~\ref{sr1}. Here, ``Th." represents the lowest mass threshold, and ``$\cdots$" means that the $S$-wave components for the corresponding channels do not exist or the corresponding channels below the threshold considered.}\label{cr1}
\begin{tabular}{c|c|cccc}\toprule[1pt]\toprule[1pt]
$J^{PC}$& Th. &$\Lambda$ &$E$&$r_{\rm RMS}$&$P(D_s\bar D_{s1}/D_s\bar D_{s2}^{*}/D_s^{*}\bar D_{s1}/D_s^{*}\bar D_{s2}^{*})$ \\\midrule[1.0pt]
\multirow{6}{*}{$1^{--}$}&$D_s\bar D_{s1}$&1.92&$-0.32$&4.42&\textbf{91.19}/$\cdots$/$o(0)$/8.18\\
             &           &1.93&$-3.38$&1.47&\textbf{72.37}/$\cdots$/$o(0)$/27.63\\
             &           &1.94&$-9.24$&0.83&\textbf{57.50}/$\cdots$/$o(0)$/\textbf{42.50}\\
         &$D_s^{*}\bar D_{s1}$&1.86&$-2.17$&0.56&$\cdots$/$\cdots$/$o(0)$/\textbf{100.00}\\
             &                &1.87&$-6.01$&0.54&$\cdots$/$\cdots$/$o(0)$/\textbf{100.00}\\
             &                &1.88&$-10.06$&0.52&$\cdots$/$\cdots$/$o(0)$/\textbf{100.00}\\
\cline{1-6}
\multirow{6}{*}{$1^{-+}$}&$D_s\bar D_{s1}$&1.87&$-0.50$&3.36&\textbf{69.27}/$\cdots$/$o(0)$/30.73\\
             &           &1.88&$-8.88$&0.61&25.86/$\cdots$/$o(0)$/\textbf{74.14}\\
             &           &1.89&$-20.56$&0.39&16.13/$\cdots$/$o(0)$/\textbf{83.87}\\
         &$D_s^{*}\bar D_{s1}$&1.74&$-0.43$&0.58&$\cdots$/$\cdots$/$o(0)$/\textbf{100.00}\\
             &                &1.75&$-5.27$&0.55&$\cdots$/$\cdots$/$o(0)$/\textbf{100.00}\\
             &                &1.76&$-10.46$&0.52&$\cdots$/$\cdots$/$o(0)$/\textbf{100.00}\\
\cline{1-6}
\multirow{6}{*}{$2^{--}$}&$D_s\bar D_{s2}^{*}$&2.27&$-3.24$&0.28&$\cdots$/$o(0)$/$o(0)$/\textbf{100.00}\\
             &               &2.28&$-13.95$&0.27&$\cdots$/$o(0)$/$o(0)$/\textbf{100.00}\\
             &               &2.29&$-25.24$&0.26&$\cdots$/$o(0)$/$o(0)$/\textbf{100.00}\\
         &$D_s^{*}\bar D_{s1}$&2.11&$-3.89$&0.54&$\cdots$/$\cdots$/$o(0)$/\textbf{100.00}\\
             &                &2.12&$-8.00$&0.51&$\cdots$/$\cdots$/$o(0)$/\textbf{100.00}\\
             &                &2.13&$-12.41$&0.49&$\cdots$/$\cdots$/$o(0)$/\textbf{100.00}\\
\cline{1-6}
\multirow{6}{*}{$2^{-+}$}&$D_s\bar D_{s2}^{*}$&2.40&$-0.33$&4.61&$\cdots$/\textbf{99.94}/0.06/$o(0)$\\
             &               &2.49&$-3.47$&1.72&$\cdots$/\textbf{99.81}/0.19/$o(0)$\\
             &               &2.57&$-10.76$&1.01&$\cdots$/\textbf{99.64}/0.36/$o(0)$\\
         &$D_s^{*}\bar D_{s1}$&2.86&$-0.31$&4.74&$\cdots$/$\cdots$/\textbf{100.00}/$o(0)$\\
             &               &3.04&$-3.07$&1.90&$\cdots$/$\cdots$/\textbf{100.00}/$o(0)$\\
             &               &3.22&$-10.18$&1.13&$\cdots$/$\cdots$/\textbf{100.00}/$o(0)$\\
\bottomrule[1pt]\bottomrule[1pt]
\end{tabular}
\end{table}

By comparing the numerical results of the single channel and coupled-channel cases, it is obvious that the bound state properties will change accordingly after considering the coupled-channel effect, i.e., the cutoff parameters in the coupled-channel analysis are smaller than that in the single channel analysis with the same binding energy in many cases, especially for the $D_s\bar D_{s1}$ state with $J^{PC}$=$1^{--}$. However, our result indicates that the coupled-channel effect to the $S$-wave $H_s \bar T_s $ systems is not obvious.

We may predict the existence of four possible hidden-charm and hidden-strange molecular states, which are the $D_s^{*}\bar D_{s1}$ states with $J^{PC}$=$0^{--}, 0^{-+}$ and the $D_s^{*}\bar D_{s2}^{*}$ states with $J^{PC}$=$1^{--}, 1^{-+}$. These predictions can be accessible at future experiment. Moreover, the $D_s^{*}\bar D_{s1}$ state with $J^{PC}$=$0^{--}$ and the $D_s^{*}\bar D_{s2}^{*}$ state with $J^{PC}$=$1^{-+}$ have the typical exotic spin-parity quantum numbers, which can be distinguished with conventional mesons.

\subsection{The open-charm and open-strange molecular systems}\label{sec1}

In the above subsection, we mainly discussed the bound state properties of the $S$-wave $H_s \bar T_s $ systems, which stimulates our interest to investigate the behavior of the open-charm and open-strange molecular systems composed by a charmed-strange meson in $H$-doublet and a charmed-strange meson in $T$-doublet, which have typical exotic state configurations totally different from the conventional hadrons.

\subsubsection{The single channel analysis}

In the following, we try to search for the bound state solutions for the $S$-wave $H_s T_s$ systems by solving the coupled-channel Schr$\ddot{\rm{o}}$dinger equation. When considering the $S$-$D$ wave mixing effect and scanning the cutoff parameters $\Lambda$ from $1$ GeV to $5~{\rm GeV}$, we list these typical values of the cutoff parameter $\Lambda$, the binding energy $E$, the root-mean-square radius $r_{\rm RMS}$, and the probability of the individual channel $P_i$ for the $S$-wave $H_s T_s$ systems in Table~\ref{sr5}. According to the theoretical results of studying the deuteron via the OBE model, the cutoff parameter $\Lambda$ should be around 1 GeV \cite{Tornqvist:1993ng,Tornqvist:1993vu,Wang:2019nwt}. If taking this criterion, our results disfavor the existence of the open-charm and open-strange molecular states for the $S$-wave $D_s^{(*)}D_{s1}( D_{s2}^{*})$ systems since the cutoff parameter $\Lambda$ is obviously far away from 1 GeV.
\renewcommand\tabcolsep{0.11cm}
\renewcommand{\arraystretch}{1.50}
\begin{table}[!htbp]
\caption{Bound state solutions for the $S$-wave $H_s T_s$ systems. Conventions are the same as Table~\ref{sr1}.}\label{sr5}
\begin{tabular}{ccc|cccc}\toprule[1pt]\toprule[1pt]
\multicolumn{7}{c}{$D_s^{*} D_{s1}\,(J^{P}=2^-)$}\\
\cline{1-7}
$\Lambda$ &$E$&$r_{\rm RMS}$&$\Lambda$ &$E$&$r_{\rm RMS}$&$P({}^5\mathbb{S}_{2}/{}^1\mathbb{D}_{2}/{}^3\mathbb{D}_{2}/{}^5\mathbb{D}_{2})$ \\
4.69&$-1.73$&1.90     &3.32&$-0.64$&3.49&\textbf{97.11}/0.13/$o(0)$/2.76\\
4.70&$-5.74$&0.98     &3.36&$-4.90$&1.33&\textbf{93.64}/0.27/$o(0)$/6.09\\
4.71&$-10.84$&0.70     &3.39&$-10.27$&0.93&\textbf{91.95}/0.33/$o(0)$/7.72\\
\cline{1-7}
\multicolumn{7}{c}{$D_s^{*} D_{s2}^{*}\,(J^{P}=2^-)$} \\
\cline{1-7}
$\Lambda$ &$E$&$r_{\rm RMS}$&$\Lambda$ &$E$&$r_{\rm RMS}$&$P({}^5\mathbb{S}_{2}/{}^3\mathbb{D}_{2}/{}^5\mathbb{D}_{2}/{}^7\mathbb{D}_{2})$ \\
$\times$&$\times$&$\times$     &4.66&$-0.25$&4.85&\textbf{98.21}/0.49/0.98/0.33\\
$\times$&$\times$&$\times$     &4.76&$-3.52$&1.62&\textbf{94.49}/1.44/3.06/1.02\\
$\times$&$\times$&$\times$     &4.86&$-10.32$&0.99&\textbf{91.66}/2.09/4.69/1.56\\
\cline{1-7}
\multicolumn{7}{c}{$D_s^{*} D_{s2}^{*}\,(J^{P}=3^-)$} \\
\cline{1-7}
$\Lambda$ &$E$&$r_{\rm RMS}$&$\Lambda$ &$E$&$r_{\rm RMS}$&$P({}^7\mathbb{S}_{3}/{}^3\mathbb{D}_{3}/{}^5\mathbb{D}_{3}/{}^7\mathbb{D}_{3})$ \\
4.20&$-2.59$&1.53     &3.14&$-0.65$&3.46&\textbf{96.94}/0.09/0.31/3.86\\
4.21&$-6.99$&0.90     &3.17&$-3.89$&1.48&\textbf{93.83}/0.13/0.44/5.60\\
4.22&$-12.40$&0.66     &3.20&$-9.38$&0.97&\textbf{91.80}/0.16/0.58/7.47\\
\bottomrule[1pt]\bottomrule[1pt]
\end{tabular}
\end{table}

\subsubsection{The coupled-channel analysis}

Similar to the $S$-wave $H_s \bar T_s$ systems, we also present the bound state properties for the $S$-wave $H_s T_s$ coupled-channel systems in Table \ref{cr2}. Nevertheless, it is clear that the coupled-channel effect plays a
positive but minor effect to generate these $S$-wave $D_s^{(*)}D_{s1}( D_{s2}^{*})$ bound states, where the corresponding cutoff is deviated from the usual value around 1 GeV for these investigated open-charm and open-strange systems \cite{Tornqvist:1993ng,Tornqvist:1993vu,Wang:2019nwt}. It means that $S$-wave bound states for the  $H_s T_s$ systems do not exist even the coupled-channel effect is included.
\renewcommand\tabcolsep{0.10cm}
\renewcommand{\arraystretch}{1.50}
\begin{table}[!htbp]
\caption{Bound state solutions for the $S$-wave $H_s T_s$ coupled-channel systems. Conventions are the same as Table~\ref{cr1}.}\label{cr2}
\begin{tabular}{c|c|cccc}\toprule[1pt]\toprule[1pt]
$J^{P}$& Th. &$\Lambda$ &$E$&$r_{\rm RMS}$&$P(D_s D_{s1}/D_s D_{s2}^{*}/D_s^{*} D_{s1}/D_s^{*}D_{s2}^{*})$ \\\midrule[1.0pt]
\multirow{6}{*}{$2^{-}$}&$D_sD_{s2}^{*}$&2.92&$-1.75$&1.62&$\cdots$/\textbf{60.58}/34.02/5.40\\
       &              &2.93&$-7.86$&0.67&$\cdots$/\textbf{47.20}/\textbf{45.48}/7.32\\
       &              &2.94&$-15.15$&0.47&$\cdots$/\textbf{42.22}/\textbf{49.65}/8.13\\
       &$D_s^{*} D_{s1}$&3.25&$-0.30$&4.51&$\cdots$/$\cdots$/\textbf{99.88}/0.12\\
       &            &3.29&$-4.45$&1.36&$\cdots$/$\cdots$/\textbf{99.58}/0.42\\
       &            &3.32&$-10.19$&0.91&$\cdots$/$\cdots$/\textbf{99.37}/0.63\\
\bottomrule[1pt]\bottomrule[1pt]
\end{tabular}
\end{table}

In short summary, since the attractive interactions between a charmed-strange meson in $H$-doublet and a charmed-strange meson in $T$-doublet are not strong enough to make the hadronic component bound together when taking reasonable input parameters. Thus, we conclude that the quantitative analysis does not support the existence of the open-charm and open-strange tetraquark molecular candidates with the $S$-wave $D_s^{(*)}D_{s1}(D_{s2}^{*})$ systems, such qualitative conclusion can be further tested in future experiment and other theoretical approaches.

\section{Summary}\label{sec4}

Exploring the exotic hadronic states has become an intriguing research issue full of challenges and opportunities, which has been inspired by the abundant observations of a series of $X/Y/Z/P_c$ states since 2003 \cite{Chen:2016qju,Liu:2019zoy}. Among different exotic hadronic configurations, hadronic molecular state has aroused heated discussion. By the efforts from both theorist and experimentalist, the properties of hidden-charm molecular states with masses below 4.5 GeV, which are from the interaction between charmed meson and anti-charmed meson, become more and more clearer.

Recently, the observation of $XYZ$ charmonium-like states in experiment collaborations \cite{Aaij:2016iza,Jia:2019gfe,Pakhlova:2008vn,Abazov:2007sf} and the announced white paper on the future physics program by the BESIII Collaboration \cite{Ablikim:2019hff} show that hunting new $XYZ$ charmonium-like states with mass above $4.5$ GeV becomes possible. Considering the close relation of $XYZ$ states with hidden-charm hadronic molecular states, we propose to perform theoretical study of hidden-charm hadronic molecular states composed of the charmed-strange meson in $H$-doublet and the anti-charmed-strange meson in $T$-doublet. The discussed hidden-charm and hidden-strange hadronic molecular states just exist in the mass range around $4.5\sim 4.7$ GeV.
We predict the existence of the $D_s^{*}\bar D_{s1}$ molecular states with $J^{PC}$=$0^{--}, 0^{-+}$ and the $D_s^{*}\bar D_{s2}^{*}$ molecular states with $J^{PC}$=$1^{--}, 1^{-+}$.
Here, the $D_s^{*}\bar D_{s1}$ molecular state with $J^{PC}$=$0^{--}$ and the $D_s^{*}\bar D_{s2}^{*}$ molecular state with $J^{PC}$=$1^{-+}$ have the exotic spin-parity quantum numbers, which can be totally distinguished with the conventional meson states.

We also extend our theoretical framework to study open-charm and open-strange molecular systems, which have components, a charmed-strange meson in $H$-doublet and a charmed-strange meson in $T$-doublet. However, our calculation does not support the existence of such hadronic molecular states.

{The possible hidden-charm hadronic molecular states with hidden strangeness can be searched in their possible two-body strong decay channels, where they can decay into a charmonium plus a $\phi$ meson, and a charmed-strange meson plus an anti-charmed-strange meson if kinetically allowed, i.e., $\eta_{c}(1S)\phi$, $J/\psi\phi$, $\chi_{cJ}(1P)\phi\,(J=0,1,2)$, $D_s \bar{D}_s$, $D_s^* \bar{D}_s$, $D_s^* \bar{D}_s^*$, and so on. These decay modes can provide vaulable information when searching for possible hidden-charm hadronic molecular states with hidden strangeness experimentally.}

Considering the experimental potential, we strongly suggest that the BESIII  Collaboration should focus on the predicted vector hidden-charm and hidden-strange molecular state by the accumulated data from $e^+e^-$ collision with $\sqrt{s}>4.5$ GeV. And, the remaining three molecular states with $J^{PC}=0^{--},0^{-+},1^{-+}$ predicted in this work can be accessible at the LHCb and BelleII by $B$ meson decays. Additionally, theoretical study of this new type of hadronic molecular state by other approaches is also encouraged. We believe that these investigations will make our knowledge of exotic hadronic molecular states become more abundant.

\section*{ACKNOWLEDGMENTS}

This work is supported by the China National Funds for Distinguished Young Scientists under Grant No. 11825503 and by the National Program for Support of Top-notch Young Professionals.

\appendix

\section{The effective Lagrangians}\label{app00}

The expanded effective Lagrangians for depicting the interactions of the (anti-)charmed mesons with the light mesons are expressed as
\begin{eqnarray}\label{eq:lagcon}
\mathcal{L}_{HH\mathbb{P}}&=&-\frac{2ig}{f_{\pi}}v^{\alpha}\varepsilon_{\alpha\mu\nu\lambda}D_b^{*\mu}D_a^{*\lambda\dag}\partial^{\nu}{\mathbb{P}}_{ba}\nonumber\\
    &&-\frac{2g}{f_{\pi}}(D_b^{*\mu}D_a^{\dag}+D_bD_a^{*\mu\dag})\partial_{\mu}{\mathbb{P}}_{ba}\nonumber\\
    &&+\frac{2ig}{f_{\pi}}v^{\alpha}\varepsilon_{\alpha\mu\nu\lambda}{\bar D}_a^{*\mu\dag}{\bar D}_b^{*\lambda}\partial^{\nu}{\mathbb{P}}_{ab}\nonumber\\
    &&+\frac{2g}{f_{\pi}}({\bar D}_a^{*\mu\dag}{\bar D}_b+{\bar D}_a^{\dag}{\bar D}_b^{*\mu})\partial_{\mu}{\mathbb{P}}_{ab},\\
\mathcal{L}_{HH\mathbb{V}} &=&-\sqrt{2}\beta g_V D_b D_a^{\dag} v\cdot\mathbb{V}_{ba}+\sqrt{2}\beta g_V D_{b\mu}^* D_a^{*\mu\dag}v\cdot\mathbb{V}_{ba}\nonumber\\
    &&-2\sqrt{2}i\lambda g_V D_b^{*\mu}D_a^{*\nu\dag}\left(\partial_{\mu}\mathbb{V}_{\nu}-\partial_{\nu}\mathbb{V}_{\mu}\right)_{ba}\nonumber\\
    &&-2\sqrt{2}\lambda g_V v^{\lambda}\varepsilon_{\lambda\mu\alpha\beta}(D_bD_a^{*\mu\dag}+D_b^{*\mu}D_a^{\dag})\partial^{\alpha}\mathbb{V}^{\beta}_{ba}\nonumber\\
    &&+\sqrt{2}\beta g_V {\bar D}_a {\bar D}_b^{\dag} v\cdot\mathbb{V}_{ab}-\sqrt{2}\beta g_V {\bar D}_{a\mu}^* {\bar D}_b^{*\mu\dag}v\cdot\mathbb{V}_{ab}\nonumber\\
    &&-2\sqrt{2}i\lambda g_V {\bar D}_a^{*\mu\dag}{\bar D}_b^{*\nu}\left(\partial_{\mu}\mathbb{V}_{\nu}-\partial_{\nu}\mathbb{V}_{\mu}\right)_{ab}\nonumber\\
    &&-2\sqrt{2}\lambda g_V v^{\lambda}\varepsilon_{\lambda\mu\alpha\beta}({\bar D}_a^{*\mu\dag}{\bar D}_b+{\bar D}_a^{\dag}{\bar D}_b^{*\mu})\partial^{\alpha}\mathbb{V}^{\beta}_{ab},\\
\mathcal {L}_{T T{\mathbb{P}}}&=&-\frac{5ik}{3f_\pi}~\epsilon^{\mu\nu\rho\tau}v_\tau D^{\dagger}_{1b\mu}D_{1a\nu} \partial_\rho\mathbb{P}_{ba}\nonumber\\
    &&+\frac{2ik}{f_\pi}~\epsilon^{\mu\nu\rho\tau}v_\nu D^{*\alpha\dagger}_{2b\rho}D^{*}_{2a\alpha\tau}\partial_\mu\mathbb{P}_{ba}\nonumber\\
    &&-\sqrt{\frac{2}{3}}\frac{k}{f_\pi}(D^{*\mu\lambda}_{2b}D^{\dagger}_{1a\mu}+D_{1b\mu}D^{*\mu\lambda\dagger}_{2a})\partial_\lambda\mathbb{P}_{ba}\nonumber\\
    &&-\frac{5ik}{3f_\pi}\varepsilon^{\mu\nu\rho\tau}v_\nu\bar{D}^{\dag}_{1a\rho}\bar{D}_{1b\tau}\partial_\mu\mathbb{P}_{ab}\nonumber\\
    &&+\frac{2ik}{f_\pi}\varepsilon^{\mu\nu\rho\tau}v_\nu\bar{D}^{*\alpha\dag}_{2a\rho}\bar{D}^{*}_{2b\alpha\tau}\partial_\mu\mathbb{P}_{ab}\nonumber\\
    &&+\sqrt{\frac{2}{3}}\frac{k}{f_\pi}\left(\bar{D}^{*\mu\lambda}_{2a}\bar{D}^{\dagger}_{1b\mu}+\bar{D}_{1a\mu}\bar{D}^{*\mu\lambda\dagger}_{2b}\right)\partial_\lambda\mathbb{P}_{ab},\\
\mathcal {L}_{TT\mathbb{V}}&=&-\sqrt{2}\beta^{\prime \prime}g_{V}(v\cdot\mathbb{V}_{ba}) D_{1b\mu}D^{\mu\dagger}_{1a}\nonumber\\
&&+\frac{5\sqrt{2}i\lambda^{\prime\prime} g_{V}}{3}(D^{\nu\dagger}_{1b}D^{\mu}_{1a}-D^{\nu}_{1b}D^{\mu\dagger}_{1a})\partial_\mu \mathbb{V}_{ba\nu}\nonumber\\
&&+\sqrt{2}\beta^{\prime \prime}g_{V}(v\cdot\mathbb{V}_{ba}) D_{2b}^{*\lambda\nu}  D^{*\dagger}_{2a{\lambda\nu}}+2\sqrt{2}i\lambda^{\prime\prime} g_{V}\nonumber\\
&&\times(D^{*\lambda\nu}_{2b} D^{*\mu\dagger}_{2a\lambda}-D^{*\lambda\nu\dagger}_{2a}D^{*\mu}_{2b\lambda} )\partial_\mu \mathbb{V}_{ba\nu}\nonumber\\
&&+\frac{i\beta^{\prime \prime}g_{V}}{\sqrt{3}}\epsilon^{\lambda\alpha\rho\tau}v_{\rho}(v\cdot
\mathbb{V}_{ba})(D^{\dagger}_{1b\alpha}D^{*}_{2a\lambda\tau}-D_{1b\alpha}
D^{\dagger*}_{2a\lambda\tau})\nonumber\\
&&+\frac{2\lambda^{\prime\prime} g_{V}}{\sqrt{3}}[3\epsilon^{\mu\lambda\nu\tau}v_\lambda(D^{\alpha\dagger}_{1a}
D^{*}_{2b\alpha\tau}+D^{\alpha}_{1b}D^{*\dagger}_{2a\alpha\tau})\partial_\mu \mathbb{V}_{ba\nu}\nonumber\\
&&+2\epsilon^{\lambda\alpha\rho\nu}v_\rho(D^{\dagger}_{1b\alpha}D^{*\mu}_{2a\lambda}
+D_{1b\alpha}D^{\dagger\mu*}_{2a\lambda})\nonumber\\
&&\times(\partial_\mu \mathbb{V}_{\nu}-\partial_\nu \mathbb{V}_{\mu})_{ba}]\nonumber\\
&&+\sqrt{2}\beta^{\prime\prime}g_{V}\left(v\cdot\mathbb{V}_{ab}\right)\bar{D}_{1a\mu}\bar{D}^{\mu\dagger}_{1b}\nonumber\\
&&+\frac{5\sqrt{2}i\lambda^{\prime\prime}g_{V}}{3}\left(\bar{D}^{\nu}_{1a}\bar{D}^{\mu\dagger}_{1b}
-\bar{D}^{\nu\dagger}_{1a}\bar{D}^{\mu}_{1b}\right)\partial_\mu\mathbb{V}_{ab\nu}\nonumber\\
&&-\sqrt{2}\beta^{\prime \prime}g_{V}\left(v\cdot\mathbb{V}_{ab}\right) \bar{D}_{2a}^{*\lambda\nu} \bar{D}^{*\dagger}_{2b{\lambda\nu}}\nonumber\\
&&+2\sqrt{2}i\lambda^{\prime\prime} g_{V}\left(\bar{D}^{*\lambda\nu\dagger}_{2a}
\bar{D}^{*\mu}_{2b\lambda}-\bar{D}^{*\lambda\nu}_{2a} \bar{D}^{*\mu\dagger}_{2b\lambda}\right)\partial_\mu \mathbb{V}_{ab\nu}\nonumber\\
&&+\frac{i\beta^{\prime\prime}g_{V}}{\sqrt{3}}\varepsilon^{\lambda\alpha\rho\tau}v_{\rho}
\left(v\cdot\mathbb{V}_{ba}\right)\left(\bar{D}^{\dagger}_{1a\alpha} \bar{D}^{*}_{2b\lambda\tau}-\bar{D}_{1a\alpha}
\bar{D}^{\dagger*}_{2b\lambda\tau}\right)\nonumber\\
&&+\frac{2\lambda^{\prime\prime}g_{V}}{\sqrt{3}}
\left[3\varepsilon^{\mu\lambda\nu\tau}v_\lambda\left(\bar{D}^{\alpha\dagger}_{1a}
\bar{D}^{*}_{2b\alpha\tau}+\bar{D}^{\alpha}_{1a}
\bar{D}^{*\dagger}_{2b\alpha\tau}\right)\partial_\mu\mathbb{V}_{ab\nu}\right.\nonumber\\
&&\left.+2\varepsilon^{\lambda\alpha\rho\nu}v_\rho
\left(\bar{D}^{\dagger}_{1a\alpha}\bar{D}^{*\mu}_{2b\lambda}
+\bar{D}_{1a\alpha}\bar{D}^{\dagger\mu*}_{2b\lambda}\right)\right.\nonumber\\
&&\left.\times\left(\partial_\mu \mathbb{V}_{\nu}-\partial_\nu \mathbb{V}_{\mu}\right)_{ab}\right],\\
\mathcal{L}_{HT{\mathbb{P}}}&=&-\frac{2h^{\prime}}{f_\pi}\left(D_bD_{2a}^{*\mu\nu\dagger}+D_{2b}^{*\mu\nu}D_a^{\dagger}\right)\partial_{\mu}\partial_{\nu}{\mathbb{P}}_{ba}\nonumber\\
    &&-\frac{\sqrt{2}h^{\prime}}{\sqrt{3}f_\pi}\left[3\left(D_{b}^{*\mu}D_{1a}^{\nu\dagger}+D_{1b}^{\nu}D_{a}^{*\mu\dagger}\right)\partial_{\mu}\partial_{\nu}{\mathbb{P}}_{ba}\right.\nonumber\\
    &&\left.+\left(D_{\lambda b}^{*}D_{1a}^{\lambda\dagger}+D_{1b}^{\lambda}D_{\lambda a}^{*\dagger}\right)(\partial_{\nu}\partial^{\nu}-v^{\mu}v^{\nu}\partial_{\mu}\partial_{\nu}){\mathbb{P}}_{ba}\right]\nonumber\\
    &&-\frac{2ih^{\prime}}{f_\pi}\varepsilon^{\lambda\nu\rho\tau}v_\lambda\left(D_{b\tau}^{*}D_{2a\rho}^{*\mu\dagger}-D_{2b\rho}^{*\mu}D_{a\tau}^{*\dagger}\right)\partial_{\mu}\partial_{\nu}{\mathbb{P}}_{ba}\nonumber\\
    &&+\frac{2h^{\prime}}{f_\pi}\left(\bar D_a\bar D_{2b}^{*\mu\nu\dagger}+\bar D_{2a}^{*\mu\nu}\bar D_b^{\dagger}\right)\partial_{\mu}\partial_{\nu}{\mathbb{P}}_{ab}\nonumber\\
    &&+\frac{\sqrt{2}h^{\prime}}{\sqrt{3}f_\pi}\left[3\left(\bar D_{a}^{*\mu}\bar D_{1b}^{\nu\dagger}+\bar D_{1a}^{\nu}\bar D_{b}^{*\mu\dagger}\right)\partial_{\mu}\partial_{\nu}{\mathbb{P}}_{ab}\right.\nonumber\\
    &&\left.+\left(\bar D_{\lambda a}^{*}\bar D_{1b}^{\lambda\dagger}+\bar D_{1a}^{\lambda}\bar D_{\lambda b}^{*\dagger}\right)(\partial_{\nu}\partial^{\nu}-v^{\mu}v^{\nu}\partial_{\mu}\partial_{\nu}){\mathbb{P}}_{ab}\right]\nonumber\\
    &&-\frac{2ih^{\prime}}{f_\pi}\varepsilon^{\lambda\nu\rho\tau}v_\lambda\left(\bar D_{2a\tau}^{*\mu\dagger}\bar D_{b\nu}^{*}-\bar D_{a\nu}^{*\dagger}\bar D_{2b\tau}^{*\mu}\right)\partial_{\mu}\partial_{\rho}{\mathbb{P}}_{ab},\\
\mathcal{L}_{HT\mathbb{V}}&=&-\frac{2}{\sqrt{3}}\zeta_1g_{V}\left(D_bD_{1a\mu}^{\dagger}+D_{1b\mu}D_a^{\dagger}\right)\mathbb{V}_{ba}^{\mu}+\frac{2i}{\sqrt{3}}\mu_1g_{V}\nonumber\\
    &&\times\left[\left(D_bD_{1a}^{\nu\dagger}-D_{1b\nu}D_a^{\dagger}\right)v^{\mu}\left(\partial_\mu \mathbb{V}_{\nu}-\partial_\nu\mathbb{V}_{\mu}\right)_{ba}\right]\nonumber\\
    &&+\frac{i}{\sqrt{3}}\zeta_1g_{V}\varepsilon^{\mu\nu\rho\tau}v_\nu\left(D_{1b\mu}^{\dagger}D_{a\tau}^{*}-D_{b\tau}^{*\dagger}D_{1a\mu}\right)\mathbb{V}_{ba\rho}\nonumber\\
    &&+\frac{\mu_1g_{V}}{\sqrt{3}}\left[\varepsilon^{\alpha\beta\rho\tau}v_\beta v^\mu\left(D_{1b\alpha}^{\dagger}D_{a\tau}^{*}+D_{b\tau}^{*\dagger}D_{1a\alpha}\right)\partial_{\mu}\mathbb{V}_{ba\rho}\right.\nonumber\\
    &&\left.-\varepsilon^{\mu\beta\lambda\tau}v_\lambda v^\rho\left(D_{1b\beta}^{\dagger}D_{a\tau}^{*}+D_{b\tau}^{*\dagger}D_{1a\beta}\right)\partial_{\mu}\mathbb{V}_{ba\rho}\right]\nonumber\\
   &&+\sqrt{2}\zeta_1g_{V}\left(D_{2b\mu\nu}^{*\dagger}D_{a}^{*\mu}+D_{b}^{*\mu\dagger}D_{2a\mu\nu}^{*}\right)\mathbb{V}_{ba}^{\nu}\nonumber\\
   &&+\sqrt{2}i\mu_1g_{V}\left[v_{\nu}\left(D_{2b\lambda\mu}^{*\dagger}D_{a}^{*\lambda}-D_{b}^{*\lambda\dagger}D_{2a\lambda\mu}^{*}\right)\right.\nonumber\\
   &&\left.+v_{\mu}\left(D_{2b\lambda\nu}^{*}D_{a}^{*\lambda\dagger}-D_{b}^{*\lambda}D_{2a\lambda\nu}^{*\dagger}\right)\right]\partial_{\mu}\mathbb{V}_{ba}^{\nu}\nonumber\\
    &&+\frac{2}{\sqrt{3}}\zeta_1g_{V}\left(\bar D_a \bar D_{1b\mu}^{\dagger}+\bar D_{1a\mu}\bar D_b^{\dagger}\right)\mathbb{V}_{ab}^{\mu}-\frac{2i}{\sqrt{3}}\mu_1g_{V}\nonumber\\
    &&\times\left[\left(\bar D_a\bar D_{1b}^{\nu\dagger}-\bar D_{1a\nu}\bar D_b^{\dagger}\right)v^{\mu}\left(\partial_\mu \mathbb{V}_{\nu}-\partial_\nu\mathbb{V}_{\mu}\right)_{ab}\right]\nonumber\\
    &&+\frac{i}{\sqrt{3}}\zeta_1g_{V}\varepsilon^{\mu\nu\rho\tau}v_\mu\left(\bar D_{1a\rho}^{\dagger}\bar D_{b\tau}^{*}-\bar D_{a\tau}^{*\dagger}\bar D_{1b\rho}\right)\mathbb{V}_{ab\nu}\nonumber\\
   &&+\frac{\mu_1g_{V}}{\sqrt{3}}\left[\varepsilon^{\alpha\beta\rho\tau}v_\alpha v^\mu\left(\bar D_{1a\beta}^{\dagger}\bar D_{b\rho}^{*}+\bar D_{a\rho}^{*\dagger}\bar D_{1b\beta}\right)\partial_{\mu}\mathbb{V}_{ab\tau}\right.\nonumber\\
    &&\left.+\varepsilon^{\mu\beta\lambda\rho}v_\beta v^\tau\left(\bar D_{1a\lambda}^{\dagger}\bar D_{b\rho}^{*}+\bar D_{a\rho}^{*\dagger}\bar D_{1b\lambda}\right)\partial_{\mu}\mathbb{V}_{ab\tau}\right]\nonumber\\
   &&-\sqrt{2}\zeta_1g_{V}\left(\bar D_{2a\mu\nu}^{*\dagger}\bar D_{b}^{*\mu}+\bar D_{a}^{*\mu\dagger}\bar D_{2b\mu\nu}^{*}\right)\mathbb{V}_{ab}^{\nu}\nonumber\\
   &&-\sqrt{2}i\mu_1g_{V}\left[v_{\nu}\left(\bar D_{2a\lambda\mu}^{*\dagger}\bar D_{b}^{*\lambda}-\bar D_{a}^{*\lambda\dagger}\bar D_{2b\lambda\mu}^{*}\right)\right.\nonumber\\
   &&\left.+v_{\mu}\left(\bar D_{2a\lambda\nu}^{*}\bar D_{b}^{*\lambda\dagger}-\bar D_{a}^{*\lambda}\bar D_{2b\lambda\nu}^{*\dagger}\right)\right]\partial_{\mu}\mathbb{V}_{ab}^{\nu}.
\end{eqnarray}

\section{The details of these obtained effective potentials}\label{app01}

Before presenting the effective potentials, we firstly focus on several typical Fourier transforms, i.e.,
\begin{eqnarray}
&&\mathcal{F}\left\{\frac{1}{{\bm{q}}^2+m^2}\left(\frac{\Lambda^2-m^2}{\Lambda^2+{\bm{q}}^2}\right)^2\right\}=Y\left(\Lambda,m,r\right),\\
&&\mathcal{F}\left\{\frac{{\bm{q}}^2}{{\bm{q}}^2+m^2}\left(\frac{\Lambda^2-m^2}{\Lambda^2+{\bm{q}}^2}\right)^2\right\}=-\mathcal{Z}Y\left(\Lambda,m,r\right),\\
&&\mathcal{F}\left\{\frac{\left({\bm{a}}\cdot{\bm{q}}\right)\left({\bm{b}}\cdot{\bm{q}}\right)}{{\bm{q}}^2+m^2}\left(\frac{\Lambda^2-m^2}{\Lambda^2+{\bm{q}}^2}\right)^2\right\}\nonumber\\
&=&-\frac{1}{3}\left({\bm{a}}\cdot{\bm{b}}\right)\mathcal{Z}Y\left(\Lambda,m,r\right)-\frac{1}{3}T\left({\bm{a}},{\bm{b}}\right)\mathcal{T}Y\left(\Lambda,m,r\right),\\
&&\mathcal{F}\left\{\frac{\left({\bm{a}}\times{\bm{q}}\right)\cdot\left({\bm{b}}\times{\bm{q}}\right)}{{\bm{q}}^2+m^2}\left(\frac{\Lambda^2-m^2}{\Lambda^2+{\bm{q}}^2}\right)^2\right\}\nonumber\\
&=&-\frac{2}{3}\left({\bm{a}}\cdot{\bm{b}}\right)\mathcal{Z}Y\left(\Lambda,m,r\right)+\frac{1}{3}T\left({\bm{a}},{\bm{b}}\right)\mathcal{T}Y\left(\Lambda,m,r\right),\\
&&\mathcal{F}\left\{\frac{\left({\bm{a}}\cdot{\bm{q}}\right)\left({\bm{b}}\cdot{\bm{q}}\right)\left({\bm{c}}\cdot{\bm{q}}\right)\left({\bm{d}}\cdot{\bm{q}}\right)}{{\bm{q}}^2+m^2}\left(\frac{\Lambda^2-m^2}{\Lambda^2+{\bm{q}}^2}\right)^2\right\}\nonumber\\
&=&\frac{1}{27}\left[\left({\bm{a}}\cdot{\bm{b}}\right)\left({\bm{c}}\cdot{\bm{d}}\right)+\left({\bm{a}}\cdot{\bm{c}}\right)\left({\bm{b}}\cdot{\bm{d}}\right)+\left({\bm{a}}\cdot{\bm{d}}\right)\left({\bm{b}}\cdot{\bm{c}}\right)\right]\nonumber\\
&&\times\mathcal{Z}\mathcal{Z}Y\left(\Lambda,m,r\right)+\frac{1}{27}[T\left({\bm{a}},{\bm{b}}\right)T\left({\bm{c}},{\bm{d}}\right)+T\left({\bm{a}},{\bm{c}}\right)T\left({\bm{b}},{\bm{d}}\right)\nonumber\\
&&+T\left({\bm{a}},{\bm{d}}\right)T\left({\bm{b}},{\bm{c}}\right)]\mathcal{T}\mathcal{T}Y\left(\Lambda,m,r\right)+\frac{1}{54}[\left({\bm{a}}\cdot{\bm{b}}\right)T\left({\bm{c}},{\bm{d}}\right)\nonumber\\
&&+\left({\bm{a}}\cdot{\bm{c}}\right)T\left({\bm{b}},{\bm{d}}\right)+\left({\bm{a}}\cdot{\bm{d}}\right)T\left({\bm{b}},{\bm{c}}\right)+\left({\bm{c}}\cdot{\bm{d}}\right)T\left({\bm{a}},{\bm{b}}\right)\nonumber\\
&&+\left({\bm{b}}\cdot{\bm{d}}\right)T\left({\bm{a}},{\bm{c}}\right)+\left({\bm{b}}\cdot{\bm{c}}\right)T\left({\bm{a}},{\bm{d}}\right)]\left\{\mathcal{T},\mathcal{Z}\right\}Y\left(\Lambda,m,r\right).\nonumber\\\label{correction terms}
\end{eqnarray}
In the above expressions, the operators are defined as $\mathcal{Z}=\frac{1}{r^2}\frac{\partial}{\partial r}r^2\frac{\partial}{\partial r}$, $\mathcal{T}=r\frac{\partial}{\partial r}\frac{1}{r}\frac{\partial}{\partial r}$, $\{\mathcal{T},\mathcal{Z}\}=\mathcal{T}\mathcal{Z}+\mathcal{Z}\mathcal{T}$, and $T\left({\bm{a}},{\bm{b}}\right)= 3\left(\hat{\bm r} \cdot {\bm a}\right)\left(\hat{\bm r} \cdot {\bm b}\right)-{\bm a} \cdot {\bm b}$.

Through the above preparation, we can write out the effective potentials in the coordinate space for all of the investigated systems, which include
\begin{itemize}
  \item $D_s\bar D_{s1} \to D_s\bar D_{s1}$ process:
\begin{eqnarray}
\mathcal{V}_{D}&=&\frac{C}{2}\frac{\mathcal{A}_1+\mathcal{A}_1^{\prime}}{2}Y_{\phi},\\
\mathcal{V}_{C}&=&\frac{E}{3}\frac{\mathcal{A}_2+\mathcal{A}_2^{\prime}}{2}Y_{\phi0}.
\end{eqnarray}

  \item $D_s\bar D_{s2}^{*} \to D_s\bar D_{s2}^{*}$ process:
\begin{eqnarray}
\mathcal{V}_{D}&=&\frac{C}{2}\frac{\mathcal{A}_3+\mathcal{A}_3^{\prime}}{2}Y_{\phi},\\
\mathcal{V}_{C}&=&\frac{2B}{3}\left[\frac{\mathcal{A}_4+\mathcal{A}_4^{\prime}}{2}\mathcal{Z}\mathcal{Z}+\frac{\mathcal{A}_5+\mathcal{A}_5^{\prime}}{2}\mathcal{T}\mathcal{T}+\frac{\mathcal{A}_6+\mathcal{A}_6^{\prime}}{2}\right.\nonumber\\
&&\left.\times\{\mathcal{T},\mathcal{Z}\}\right]Y_{\eta1}.
\end{eqnarray}

  \item $D_s^{*}\bar D_{s1} \to D_s^{*}\bar D_{s1}$ process:
\begin{eqnarray}
\mathcal{V}_{D}&=&\frac{5A}{27}\left[\mathcal{A}_8\mathcal{Z}+\mathcal{A}_9\mathcal{T}\right]Y_{\eta}\nonumber\\
&&+\left[\frac{C}{2}\mathcal{A}_7+\frac{5D}{9}\left(\mathcal{A}_9\mathcal{T}-2\mathcal{A}_8\mathcal{Z}\right)\right]Y_{\phi},\\
\mathcal{V}_{C}&=&\frac{B}{9}\left[\mathcal{A}_{10}\mathcal{Z}\mathcal{Z}+\mathcal{A}_{11}\mathcal{T}\mathcal{T}+\mathcal{A}_{12}\{\mathcal{T},\mathcal{Z}\}\right]Y_{\eta2}\nonumber\\
&&+\frac{E}{12}\mathcal{A}_{8}Y_{\phi2}.
\end{eqnarray}

  \item $D_s^{*}\bar D_{s2}^{*} \to D_s^{*}\bar D_{s2}^{*}$ process:
\begin{eqnarray}
\mathcal{V}_{D}&=&\frac{2A}{9}\left[\frac{\mathcal{A}_{14}+\mathcal{A}_{14}^{\prime}}{2}\mathcal{Z}+\frac{\mathcal{A}_{15}+\mathcal{A}_{15}^{\prime}}{2}\mathcal{T}\right]Y_{\eta}\nonumber\\
&&+\left[\frac{2D}{3}\left(\frac{\mathcal{A}_{15}+\mathcal{A}_{15}^{\prime}}{2}\mathcal{T}-2\frac{\mathcal{A}_{14}+\mathcal{A}_{14}^{\prime}}{2}\mathcal{Z}\right)\right.\nonumber\\
&&+\left.\frac{C}{2}\frac{\mathcal{A}_{13}+\mathcal{A}_{13}^{\prime}}{2}\right]Y_{\phi},\\
\mathcal{V}_{C}&=&\frac{2B}{3}\left[\frac{\mathcal{A}_{16}+\mathcal{A}_{16}^{\prime}}{2}\mathcal{Z}\mathcal{Z}+\frac{\mathcal{A}_{17}+\mathcal{A}_{17}^{\prime}}{2}\mathcal{T}\mathcal{T}\right.\nonumber\\
&&\left.+\frac{\mathcal{A}_{18}+\mathcal{A}_{18}^{\prime}}{2}\{\mathcal{T},\mathcal{Z}\}\right]Y_{\eta3}+\frac{E}{2}\frac{\mathcal{A}_{19}+\mathcal{A}_{19}^{\prime}}{2}Y_{\phi3}.\nonumber\\
\end{eqnarray}
\end{itemize}
Here, $A=g k/f_\pi^2$, $B=h^{\prime2}/f_\pi^2$, $C=\beta \beta^{\prime\prime} g_{V}^2$, $D= \lambda \lambda^{\prime\prime}g_V^2$, and $E=\zeta_1^2g_{V}^2$. In the above expressions, the function $Y(\Lambda_i, m_i, r)$ reads as
\begin{eqnarray}
Y_i\equiv Y(\Lambda_i, m_i, r)=\dfrac{e^{-m_ir}-e^{-\Lambda_ir}}{4\pi r}-\dfrac{\Lambda_i^2-m_i^2}{8\pi\Lambda_i}e^{-\Lambda_ir}
\end{eqnarray}
with $m_i=\sqrt{m^2-q_i^2}$ and $\Lambda_i=\sqrt{\Lambda^2-q_i^2}$. The variables $q_i$ are defined as $q_0 = m_{D_{s1}}-m_{D_s}$, $q_1 = m_{D_{s2}^{*}}-m_{D_s}$, $q_2 =m_{D_{s1}}-m_{D_s^*},$ and $q_3= m_{D_{s2}^{*}}-m_{D_s^*}$. In addition, we introduce several operators, which include
\begin{eqnarray}\label{op}
\mathcal{A}_{1}&=&{\bm\epsilon^{\dagger}_4}\cdot{\bm\epsilon_2},~~~\mathcal{A}_{1}^{\prime}={\bm\epsilon^{\dagger}_3}\cdot{\bm\epsilon_1},~~~\mathcal{A}_{2}={\bm\epsilon^{\dagger}_3}\cdot{\bm\epsilon_2},~~~\mathcal{A}_{2}^{\prime}={\bm\epsilon^{\dagger}_4}\cdot{\bm\epsilon_1},\nonumber\\
\mathcal{A}_{3}&=&\mathcal{\sum}\left({\bm\epsilon^{\dagger}_{4m}}\cdot{\bm\epsilon_{2a}}\right)\left({\bm\epsilon^{\dagger}_{4n}}\cdot {\bm\epsilon_{2b}}\right),\nonumber\\
\mathcal{A}_{3}^{\prime}&=&\mathcal{\sum}\left({\bm\epsilon^{\dagger}_{3m}}\cdot{\bm\epsilon_{1a}}\right)\left({\bm\epsilon^{\dagger}_{3n}}\cdot {\bm\epsilon_{1b}}\right),\nonumber\\
\mathcal{A}_{4}&=&\frac{2}{27}\mathcal{\sum}\left({\bm\epsilon^{\dagger}_{3m}}\cdot{\bm\epsilon_{2a}}\right)\left({\bm\epsilon^{\dagger}_{3n}}\cdot {\bm\epsilon_{2b}}\right),\nonumber\\
\mathcal{A}_{4}^{\prime}&=&\frac{2}{27}\mathcal{\sum}\left({\bm\epsilon^{\dagger}_{4m}}\cdot{\bm\epsilon_{1a}}\right)\left({\bm\epsilon^{\dagger}_{4n}}\cdot {\bm\epsilon_{1b}}\right),\nonumber\\
\mathcal{A}_{5}&=&\frac{1}{27}\mathcal{\sum}T({\bm\epsilon^{\dagger}_{3m}},{\bm\epsilon^{\dagger}_{3n}})T({\bm\epsilon_{2a}},{\bm\epsilon_{2b}})\nonumber\\
&&+\frac{2}{27}\mathcal{\sum}T({\bm\epsilon^{\dagger}_{3m}},{\bm\epsilon_{2a}})T({\bm\epsilon^{\dagger}_{3n}},{\bm\epsilon_{2b}}),\nonumber\\
\mathcal{A}_{5}^{\prime}&=&\frac{1}{27}\mathcal{\sum}T({\bm\epsilon^{\dagger}_{4m}},{\bm\epsilon^{\dagger}_{4n}})T({\bm\epsilon_{1a}},{\bm\epsilon_{1b}})\nonumber\\
&&+\frac{2}{27}\mathcal{\sum}T({\bm\epsilon^{\dagger}_{4m}},{\bm\epsilon_{1a}})T({\bm\epsilon^{\dagger}_{4n}},{\bm\epsilon_{1b}}),\nonumber\\
\mathcal{A}_{6}&=&\frac{2}{27}\mathcal{\sum}\left({\bm\epsilon^{\dagger}_{3m}}\cdot{\bm\epsilon_{2a}}\right)T({\bm\epsilon^{\dagger}_{3n}},{\bm\epsilon_{2b}}),\nonumber\\
\mathcal{A}_{6}^{\prime}&=&\frac{2}{27}\mathcal{\sum}\left({\bm\epsilon^{\dagger}_{4m}}\cdot{\bm\epsilon_{1a}}\right)T({\bm\epsilon^{\dagger}_{4n}},{\bm\epsilon_{1b}}),\nonumber\\
\mathcal{A}_{7}&=&\left({\bm\epsilon^{\dagger}_3}\cdot{\bm\epsilon_1}\right)\left({\bm\epsilon^{\dagger}_4}\cdot{\bm\epsilon_2}\right),~~~~~~\mathcal{A}_{8}=\left({\bm\epsilon^{\dagger}_3}\times{\bm\epsilon_1}\right)\cdot\left({\bm\epsilon^{\dagger}_4}\times{\bm\epsilon_2}\right),\nonumber\\
\mathcal{A}_{9}&=&T({\bm\epsilon^{\dagger}_3}\times{\bm\epsilon_1},{\bm\epsilon^{\dagger}_4}\times{\bm\epsilon_2}),\nonumber\\
\mathcal{A}_{10}&=&-\frac{1}{3}\left({\bm\epsilon^{\dagger}_3}\cdot{\bm\epsilon_1}\right)\left({\bm\epsilon^{\dagger}_4}\cdot{\bm\epsilon_2}\right)
+\frac{1}{3}\left({\bm\epsilon^{\dagger}_3}\cdot{\bm\epsilon^{\dagger}_4}\right)\left({\bm\epsilon_1}\cdot{\bm\epsilon_2}\right),\nonumber\\
\mathcal{A}_{11}&=&\frac{2}{3}T({\bm\epsilon^{\dagger}_3},{\bm\epsilon_1})T({\bm\epsilon^{\dagger}_4},{\bm\epsilon_2})+\frac{1}{3}T({\bm\epsilon^{\dagger}_3},{\bm\epsilon^{\dagger}_4})T({\bm\epsilon_1},{\bm\epsilon_2}),\nonumber\\
\mathcal{A}_{12}&=&\frac{1}{6}\left({\bm\epsilon^{\dagger}_3}\cdot{\bm\epsilon^{\dagger}_4}\right)T({\bm\epsilon_1},{\bm\epsilon_2})+\frac{1}{6}\left({\bm\epsilon_1}\cdot{\bm\epsilon_2}\right)T({\bm\epsilon^{\dagger}_3},{\bm\epsilon^{\dagger}_4})\nonumber\\
&&-\frac{1}{3}\left({\bm\epsilon^{\dagger}_3}\cdot{\bm\epsilon_1}\right)T({\bm\epsilon^{\dagger}_4},{\bm\epsilon_2}),\nonumber\\
\mathcal{A}_{13}&=&\mathcal{\sum}\left({\bm\epsilon^{\dagger}_3}\cdot{\bm\epsilon_1}\right)\left({\bm\epsilon^{\dagger}_{4m}}\cdot{\bm\epsilon_{2a}}\right)\left({\bm\epsilon^{\dagger}_{4n}}\cdot {\bm\epsilon_{2b}}\right),\nonumber\\
\mathcal{A}_{13}^{\prime}&=&\mathcal{\sum}\left({\bm\epsilon^{\dagger}_4}\cdot{\bm\epsilon_2}\right)\left({\bm\epsilon^{\dagger}_{3m}}\cdot{\bm\epsilon_{1a}}\right)\left({\bm\epsilon^{\dagger}_{3n}}\cdot {\bm\epsilon_{1b}}\right),\nonumber\\
\mathcal{A}_{14}&=&\mathcal{\sum}\left({\bm\epsilon^{\dagger}_{4m}}\cdot{\bm\epsilon_{2a}}\right)\left[\left({\bm\epsilon^{\dagger}_{3}}\times{\bm\epsilon_{1}}\right)\cdot\left({\bm\epsilon^{\dagger}_{4n}}\times {\bm\epsilon_{2b}}\right)\right],\nonumber\\
\mathcal{A}_{14}^{\prime}&=&\mathcal{\sum}\left({\bm\epsilon^{\dagger}_{3m}}\cdot{\bm\epsilon_{1a}}\right)\left[\left({\bm\epsilon^{\dagger}_{4}}\times{\bm\epsilon_{2}}\right)\cdot\left({\bm\epsilon^{\dagger}_{3n}}\times {\bm\epsilon_{1b}}\right)\right],\nonumber\\
\mathcal{A}_{15}&=&\mathcal{\sum}\left({\bm\epsilon^{\dagger}_{4m}}\cdot{\bm\epsilon_{2a}}\right)T({\bm\epsilon^{\dagger}_{3}}\times{\bm\epsilon_{1}},{\bm\epsilon^{\dagger}_{4n}}\times {\bm\epsilon_{2b}}),\nonumber\\
\mathcal{A}_{15}^{\prime}&=&\mathcal{\sum}\left({\bm\epsilon^{\dagger}_{3m}}\cdot{\bm\epsilon_{1a}}\right)T({\bm\epsilon^{\dagger}_{4}}\times{\bm\epsilon_{2}},{\bm\epsilon^{\dagger}_{3n}}\times {\bm\epsilon_{1b}}),\nonumber\\
\mathcal{A}_{16}&=&\frac{1}{27}\mathcal{\sum}\left[\left({\bm\epsilon^{\dagger}_{3m}}\times{\bm\epsilon_{1}}\right)\cdot\left({\bm\epsilon^{\dagger}_{4}}\times{\bm\epsilon_{2a}}\right)\right]\left({\bm\epsilon^{\dagger}_{3n}}\cdot{\bm\epsilon_{2b}}\right)\nonumber\\
&&+\frac{1}{27}\mathcal{\sum}\left[\left({\bm\epsilon^{\dagger}_{3m}}\times{\bm\epsilon_{1}}\right)\cdot{\bm\epsilon_{2b}}\right]\left[{\bm\epsilon^{\dagger}_{3n}}\cdot\left({\bm\epsilon^{\dagger}_{4}}\times{\bm\epsilon_{2a}}\right)\right],\nonumber\\
\mathcal{A}_{16}^{\prime}&=&\frac{1}{27}\mathcal{\sum}\left[\left({\bm\epsilon^{\dagger}_{4m}}\times{\bm\epsilon_{2}}\right)\cdot\left({\bm\epsilon^{\dagger}_{3}}\times{\bm\epsilon_{1a}}\right)\right]\left({\bm\epsilon^{\dagger}_{4n}}\cdot{\bm\epsilon_{1b}}\right)\nonumber\\
&&+\frac{1}{27}\mathcal{\sum}\left[\left({\bm\epsilon^{\dagger}_{4m}}\times{\bm\epsilon_{2}}\right)\cdot{\bm\epsilon_{1b}}\right]\left[{\bm\epsilon^{\dagger}_{4n}}\cdot\left({\bm\epsilon^{\dagger}_{3}}\times{\bm\epsilon_{1a}}\right)\right],\nonumber\\
\mathcal{A}_{17}&=&\frac{1}{27}\mathcal{\sum}T({\bm\epsilon^{\dagger}_{3m}}\times{\bm\epsilon_{1}},{\bm\epsilon^{\dagger}_{3n}})T({\bm\epsilon^{\dagger}_{4}}\times{\bm\epsilon_{2a}},{\bm\epsilon_{2b}})\nonumber\\
&&+\frac{1}{27}\mathcal{\sum}T({\bm\epsilon^{\dagger}_{3m}}\times{\bm\epsilon_{1}},{\bm\epsilon^{\dagger}_{4}}\times{\bm\epsilon_{2a}})T({\bm\epsilon^{\dagger}_{3n}},{\bm\epsilon_{2b}})\nonumber\\
&&+\frac{1}{27}\mathcal{\sum}T({\bm\epsilon^{\dagger}_{3m}}\times{\bm\epsilon_{1}},{\bm\epsilon_{2b}})T({\bm\epsilon^{\dagger}_{3n}},{\bm\epsilon^{\dagger}_{4}}\times{\bm\epsilon_{2a}}),\nonumber\\
\mathcal{A}_{17}^{\prime}&=&\frac{1}{27}\mathcal{\sum}T({\bm\epsilon^{\dagger}_{4m}}\times{\bm\epsilon_{2}},{\bm\epsilon^{\dagger}_{4n}})T({\bm\epsilon^{\dagger}_{3}}\times{\bm\epsilon_{1a}},{\bm\epsilon_{1b}})\nonumber\\
&&+\frac{1}{27}\mathcal{\sum}T({\bm\epsilon^{\dagger}_{4m}}\times{\bm\epsilon_{2}},{\bm\epsilon^{\dagger}_{3}}\times{\bm\epsilon_{1a}})T({\bm\epsilon^{\dagger}_{4n}},{\bm\epsilon_{1b}})\nonumber\\
&&+\frac{1}{27}\mathcal{\sum}T({\bm\epsilon^{\dagger}_{4m}}\times{\bm\epsilon_{2}},{\bm\epsilon_{1b}})T({\bm\epsilon^{\dagger}_{4n}},{\bm\epsilon^{\dagger}_{3}}\times{\bm\epsilon_{1a}}),\nonumber\\
\mathcal{A}_{18}&=&\frac{1}{54}\mathcal{\sum}\left[\left({\bm\epsilon^{\dagger}_{3m}}\times{\bm\epsilon_{1}}\right)\cdot\left({\bm\epsilon^{\dagger}_{4}}\times{\bm\epsilon_{2a}}\right)\right]T({\bm\epsilon^{\dagger}_{3n}},{\bm\epsilon_{2b}})\nonumber\\
&&+\frac{1}{54}\mathcal{\sum}\left[\left({\bm\epsilon^{\dagger}_{3m}}\times{\bm\epsilon_{1}}\right)\cdot{\bm\epsilon_{2b}}\right]T({\bm\epsilon^{\dagger}_{3n}},{\bm\epsilon^{\dagger}_{4}}\times{\bm\epsilon_{2a}})\nonumber\\
&&+\frac{1}{54}\mathcal{\sum}\left({\bm\epsilon^{\dagger}_{3n}}\cdot{\bm\epsilon_{2b}}\right)T({\bm\epsilon^{\dagger}_{3m}}\times{\bm\epsilon_{1}},{\bm\epsilon^{\dagger}_{4}}\times{\bm\epsilon_{2a}})\nonumber\\
&&+\frac{1}{54}\mathcal{\sum}\left[{\bm\epsilon^{\dagger}_{3n}}\cdot\left({\bm\epsilon^{\dagger}_{4}}\times{\bm\epsilon_{2a}}\right)\right]T({\bm\epsilon^{\dagger}_{3m}}\times{\bm\epsilon_{1}},{\bm\epsilon_{2b}}),\nonumber\\
\mathcal{A}_{18}^{\prime}&=&\frac{1}{54}\mathcal{\sum}\left[\left({\bm\epsilon^{\dagger}_{4m}}\times{\bm\epsilon_{2}}\right)\cdot\left({\bm\epsilon^{\dagger}_{3}}\times{\bm\epsilon_{1a}}\right)\right]T({\bm\epsilon^{\dagger}_{4n}},{\bm\epsilon_{1b}})\nonumber\\
&&+\frac{1}{54}\mathcal{\sum}\left[\left({\bm\epsilon^{\dagger}_{4m}}\times{\bm\epsilon_{2}}\right)\cdot{\bm\epsilon_{1b}}\right]T({\bm\epsilon^{\dagger}_{4n}},{\bm\epsilon^{\dagger}_{3}}\times{\bm\epsilon_{1a}})\nonumber\\
&&+\frac{1}{54}\mathcal{\sum}\left({\bm\epsilon^{\dagger}_{4n}}\cdot{\bm\epsilon_{1b}}\right)T({\bm\epsilon^{\dagger}_{4m}}\times{\bm\epsilon_{2}},{\bm\epsilon^{\dagger}_{3}}\times{\bm\epsilon_{1a}})\nonumber\\
&&+\frac{1}{54}\mathcal{\sum}\left[{\bm\epsilon^{\dagger}_{4n}}\cdot\left({\bm\epsilon^{\dagger}_{3}}\times{\bm\epsilon_{1a}}\right)\right]T({\bm\epsilon^{\dagger}_{4m}}\times{\bm\epsilon_{2}},{\bm\epsilon_{1b}}),\nonumber\\
\mathcal{A}_{19}&=&\mathcal{\sum}\left({\bm\epsilon^{\dagger}_{3m}}\cdot{\bm\epsilon_1}\right)\left({\bm\epsilon^{\dagger}_4}\cdot{\bm\epsilon_{2a}}\right)\left({\bm\epsilon^{\dagger}_{3n}}\cdot {\bm\epsilon_{2b}}\right),\nonumber\\
\mathcal{A}_{19}^{\prime}&=&\mathcal{\sum}\left({\bm\epsilon^{\dagger}_{4m}}\cdot{\bm\epsilon_2}\right)\left({\bm\epsilon^{\dagger}_3}\cdot{\bm\epsilon_{1a}}\right)\left({\bm\epsilon^{\dagger}_{4n}}\cdot {\bm\epsilon_{1b}}\right).
\end{eqnarray}
Here, we define $\mathcal{\sum}=\sum_{m,n,a,b}C^{2,m+n}_{1m,1n}C^{2,a+b}_{1a,1b}$. For these operators $\mathcal{A}_k^{(\prime)}$, they should be sandwiched by the spin-orbital wave functions $|{}^{2S+1}L_{J}\rangle$, we present the relevant operator matrix elements $\mathcal{A}_k^{(\prime)}[J]$ in Table~\ref{matrix}.
\renewcommand\tabcolsep{0.00cm}
\renewcommand{\arraystretch}{1.60}
\begin{table*}[!htbp]
\caption{The relevant operator matrix elements $\mathcal{A}_k^{(\prime)}[J]\,(k=1,\cdot\cdot\cdot,19)$ for the $S$-wave $H_s \bar{T_s}$ systems.\label{matrix}}
\begin{tabular}{l|l|l}\toprule[1.5pt]
\multicolumn{3}{c}{$\mathcal{A}_k^{(\prime)}[J]=\langle f|\mathcal{A}_k^{(\prime)}|i\rangle$}\\\midrule[1.0pt]
$\begin{array}{l}\mathcal{A}_1^{(\prime)}[1]=\rm {diag}(1,1)~~~ \mathcal{A}_2^{(\prime)}[1]=\rm {diag}(1,1) \\ \mathcal{A}_3^{(\prime)}[2]=\rm {diag}(1,1)~~~ \mathcal{A}_4^{(\prime)}[2]=\rm {diag}(\frac{2}{27},\frac{2}{27})\end{array}$ &$\mathcal{A}_5^{(\prime)}[2]=\left(\begin{array}{cc} \frac{8}{135} & -\frac{4\sqrt{2}}{27\sqrt{35}} \\ -\frac{4\sqrt{2}}{27\sqrt{35}} & \frac{4}{27}\end{array}\right)$ &$\mathcal{A}_6^{(\prime)}[2]=\left(\begin{array}{cc} 0 & -\frac{\sqrt{7}}{27\sqrt{10}} \\ -\frac{\sqrt{7}}{27\sqrt{10}} & -\frac{1}{126}\end{array}\right)$\\
$\begin{array}{l}\mathcal{A}_7[0]=\rm {diag}(1,1) ~~~ \mathcal{A}_8[0]=\rm {diag}(2,-1) \\ \mathcal{A}_{10}[0]=\rm {diag}(\frac{2}{3},-\frac{1}{3})\end{array}$ & $\mathcal{A}_{9}[0]=\left(\begin{array}{cc} 0 & \sqrt{2} \\ \sqrt{2} & 2\end{array}\right)$ &$\mathcal{A}_{11}[0]=\left(\begin{array}{cc} \frac{4}{3} & -\frac{2\sqrt{2}}{3} \\ -\frac{2\sqrt{2}}{3} & 4\end{array}\right)$\\
$\mathcal{A}_{12}[0]=\left(\begin{array}{cc} 0 & \frac{\sqrt{2}}{15} \\ -\frac{8\sqrt{2}}{15} & -\frac{1}{15}\end{array}\right)$& $\begin{array}{l} \mathcal{A}_7[1]=\rm {diag}(1,1,1) \\ \mathcal{A}_8[1]=\rm {diag}(1,1,-1) \\ \mathcal{A}_{10}[1]=\rm {diag}(-\frac{1}{3},-\frac{1}{3},-\frac{1}{3}) \end{array}$&$\mathcal{A}_{9}[1]=\left(\begin{array}{ccc} 0 & -\sqrt{2} &0 \\ -\sqrt{2} & 1 &0 \\ 0&0&1\end{array}\right)$\\
$\mathcal{A}_{11}[1]=\left(\begin{array}{ccc} -\frac{2}{3} & -\frac{2\sqrt{2}}{3} &0 \\ -\frac{2\sqrt{2}}{3} & 0 &0 \\ 0&0&-\frac{4}{3}\end{array}\right)$& $\mathcal{A}_{12}[1]=\left(\begin{array}{ccc} 0 & -\frac{1}{30\sqrt{2}}&\frac{\sqrt{3}}{10\sqrt{2}} \\ -\frac{1}{30\sqrt{2}}&\frac{4}{105} &\frac{3\sqrt{3}}{70} \\ \frac{\sqrt{3}}{10\sqrt{2}}&\frac{3\sqrt{3}}{70}&-\frac{1}{105}\end{array}\right)$&$\begin{array}{l} \mathcal{A}_7[2]=\rm {diag}(1,1,1,1) \\ \mathcal{A}_8[2]=\rm {diag}(-1,2,1,-1)\\ \mathcal{A}_{10}[2]=\rm {diag}(-\frac{1}{3},\frac{2}{3},-\frac{1}{3},-\frac{1}{3}) \end{array}$ \\
$\mathcal{A}_{9}[2]=\left(\begin{array}{cccc} 0 & \frac{\sqrt{2}}{\sqrt{5}} & 0 &-\frac{\sqrt{14}}{\sqrt{5}} \\ \frac{\sqrt{2}}{\sqrt{5}} & 0 & 0 &-\frac{2}{\sqrt{7}} \\ 0 & 0 &-1 &0 \\ -\frac{\sqrt{14}}{\sqrt{5}}&-\frac{2}{\sqrt{7}}&0&-\frac{3}{7}\end{array}\right)$& $\mathcal{A}_{11}[2]=\left(\begin{array}{cccc} \frac{8}{15} &-\frac{2\sqrt{2}}{3\sqrt{5}} & 0 &-\frac{4\sqrt{2}}{3\sqrt{35}} \\-\frac{2\sqrt{2}}{3\sqrt{5}} &\frac{4}{3} & 0 &\frac{4}{3\sqrt{7}} \\ 0 &0 & -\frac{4}{3} &0 \\ -\frac{4\sqrt{2}}{3\sqrt{35}}&\frac{4}{3\sqrt{7}}&0&\frac{4}{3}\end{array}\right)$&$\mathcal{A}_{12}[2]=\left(\begin{array}{cccc} 0 &-\frac{1}{\sqrt{10}}& 0&\frac{\sqrt{2}}{3\sqrt{35}} \\ 0 &0& 0&-\frac{4}{21\sqrt{7}} \\0&0&-\frac{1}{14} &\frac{1}{14\sqrt{7}} \\ \frac{\sqrt{2}}{3\sqrt{35}}&\frac{23}{21\sqrt{7}}&\frac{1}{14\sqrt{7}}&\frac{13}{294}\end{array}\right)$\\
$\begin{array}{l} \mathcal{A}_{13}^{(\prime)}[1]=\rm {diag}(1,1,1,1) \\ \mathcal{A}_{14}^{(\prime)}[1]=\rm {diag}(\frac{3}{2},\frac{3}{2},\frac{1}{2},-1)\\\mathcal{A}_{16}^{(\prime)}[1]=\rm {diag}(\frac{1}{18},\frac{1}{18},\frac{5}{54},-\frac{1}{27})\\\mathcal{A}_{19}^{(\prime)}[1]=\rm {diag}(\frac{1}{6},\frac{1}{6},\frac{1}{2},1)\end{array}$&$\mathcal{A}_{15}[1]=\left(\begin{array}{cccc} 0&\frac{3}{5\sqrt{2}}&\frac{\sqrt{6}}{\sqrt{5}}&\frac{\sqrt{21}}{5\sqrt{2}} \\ \frac{3}{5\sqrt{2}}&-\frac{3}{10}&\frac{\sqrt{3}}{\sqrt{5}}&-\frac{\sqrt{3}}{5\sqrt{7}}\\\frac{\sqrt{6}}{\sqrt{5}}&\frac{\sqrt{3}}{\sqrt{5}}&\frac{1}{2}&\frac{2}{\sqrt{35}}\\\frac{\sqrt{21}}{5\sqrt{2}}
&-\frac{\sqrt{3}}{5\sqrt{7}}&\frac{2}{\sqrt{35}}&\frac{48}{35}\end{array}\right)$&
$\mathcal{A}_{15}^{\prime}[1]=\left(\begin{array}{cccc} 0&\frac{3}{5\sqrt{2}}&-\frac{\sqrt{6}}{\sqrt{5}}&\frac{\sqrt{21}}{5\sqrt{2}} \\ \frac{3}{5\sqrt{2}}&-\frac{3}{10}&-\frac{\sqrt{3}}{\sqrt{5}}&-\frac{\sqrt{3}}{5\sqrt{7}}\\-\frac{\sqrt{6}}{\sqrt{5}}&-\frac{\sqrt{3}}{\sqrt{5}}&\frac{1}{2}&-\frac{2}{\sqrt{35}}\\\frac{\sqrt{21}}{5\sqrt{2}}
&-\frac{\sqrt{3}}{5\sqrt{7}}&-\frac{2}{\sqrt{35}}&\frac{48}{35}\end{array}\right)$\\
$\mathcal{A}_{17}[1]=\left(\begin{array}{cccc} \frac{2}{45}&-\frac{\sqrt{2}}{45}&0&\frac{\sqrt{2}}{15\sqrt{21}} \\ -\frac{\sqrt{2}}{45}&\frac{1}{15}&0&-\frac{8}{15\sqrt{21}}\\0&0&-\frac{1}{9}&-\frac{2\sqrt{5}}{27\sqrt{7}}\\\frac{\sqrt{2}}{15\sqrt{21}}
&-\frac{8}{15\sqrt{21}}&\frac{2\sqrt{5}}{27\sqrt{7}}&\frac{92}{945}\end{array}\right)$&$\mathcal{A}_{17}^{\prime}[1]=\left(\begin{array}{cccc} \frac{2}{45}&-\frac{\sqrt{2}}{45}&0&\frac{\sqrt{2}}{15\sqrt{21}} \\ -\frac{\sqrt{2}}{45}&\frac{1}{15}&0&-\frac{8}{15\sqrt{21}}\\0&0&-\frac{1}{9}&\frac{2\sqrt{5}}{27\sqrt{7}}\\\frac{\sqrt{2}}{15\sqrt{21}}
&-\frac{8}{15\sqrt{21}}&-\frac{2\sqrt{5}}{27\sqrt{7}}&\frac{92}{945}\end{array}\right)$&
$\mathcal{A}_{18}[1]=\left(\begin{array}{cccc} 0&-\frac{7}{90\sqrt{2}}&0&\frac{\sqrt{7}}{30\sqrt{6}}\\ -\frac{7}{90\sqrt{2}}&\frac{7}{180}&0&-\frac{1}{30\sqrt{21}}\\0&0&\frac{1}{108}&\frac{\sqrt{5}}{27\sqrt{7}}\\\frac{\sqrt{7}}{30\sqrt{6}}&-\frac{1}{30\sqrt{21}}&-\frac{\sqrt{5}}{27\sqrt{7}}&\frac{4}{315}\end{array}\right)$\\
$\mathcal{A}_{18}^{\prime}[1]=\left(\begin{array}{cccc} 0&-\frac{7}{90\sqrt{2}}&0&\frac{\sqrt{7}}{30\sqrt{6}}\\ -\frac{7}{90\sqrt{2}}&\frac{7}{180}&0&-\frac{1}{30\sqrt{21}}\\0&0&\frac{1}{108}&-\frac{\sqrt{5}}{27\sqrt{7}}\\\frac{\sqrt{7}}{30\sqrt{6}}&-\frac{1}{30\sqrt{21}}&\frac{\sqrt{5}}{27\sqrt{7}}&\frac{4}{315}\end{array}\right)$ &
$\begin{array}{l} \mathcal{A}_{13}^{(\prime)}[2]=\rm {diag}(1,1,1,1) \\ \mathcal{A}_{14}^{(\prime)}[2]=\rm {diag}(\frac{1}{2},\frac{3}{2},\frac{1}{2},-1)\\\mathcal{A}_{16}^{(\prime)}[2]=\rm {diag}(\frac{5}{54},\frac{1}{18},\frac{5}{54},-\frac{1}{27})\\ \mathcal{A}_{19}^{(\prime)}[2]=\rm {diag}(\frac{1}{2},\frac{1}{6},\frac{1}{2},1) \end{array}$&$\mathcal{A}_{15}[2]=\left(\begin{array}{cccc} 0&-\frac{3\sqrt{2}}{5}&-\frac{\sqrt{7}}{\sqrt{10}}&\frac{\sqrt{7}}{5} \\ -\frac{3\sqrt{2}}{5}&\frac{3}{10}&\frac{3}{\sqrt{35}}&-\frac{3\sqrt{2}}{5\sqrt{7}}\\-\frac{\sqrt{7}}{\sqrt{10}}&\frac{3}{\sqrt{35}}&-\frac{3}{14}&\frac{4\sqrt{2}}{7\sqrt{5}}\\\frac{\sqrt{7}}{5}
&-\frac{3\sqrt{2}}{5\sqrt{7}}&\frac{4\sqrt{2}}{7\sqrt{5}}&\frac{12}{35}\end{array}\right)$\\
$\mathcal{A}_{15}^{\prime}[2]=\left(\begin{array}{cccc} 0&\frac{3\sqrt{2}}{5}&-\frac{\sqrt{7}}{\sqrt{10}}&-\frac{\sqrt{7}}{5} \\ \frac{3\sqrt{2}}{5}&\frac{3}{10}&-\frac{3}{\sqrt{35}}&-\frac{3\sqrt{2}}{5\sqrt{7}}\\-\frac{\sqrt{7}}{\sqrt{10}}&-\frac{3}{\sqrt{35}}&-\frac{3}{14}&-\frac{4\sqrt{2}}{7\sqrt{5}}\\-\frac{\sqrt{7}}{5}
&-\frac{3\sqrt{2}}{5\sqrt{7}}&-\frac{4\sqrt{2}}{7\sqrt{5}}&\frac{12}{35}\end{array}\right)$&$\mathcal{A}_{17}[2]=\left(\begin{array}{cccc} \frac{2}{27}&0&-\frac{\sqrt{2}}{27\sqrt{35}}&\frac{2}{27\sqrt{7}} \\ 0&\frac{1}{45}&0&\frac{\sqrt{2}}{15\sqrt{7}}\\-\frac{\sqrt{2}}{27\sqrt{35}}&0&\frac{29}{189}&\frac{5\sqrt{10}}{189}\\-\frac{2}{27\sqrt{7}}
&\frac{\sqrt{2}}{15\sqrt{7}}&-\frac{5\sqrt{10}}{189}&-\frac{28}{135}\end{array}\right)$&$\mathcal{A}_{17}^{\prime}[2]=\left(\begin{array}{cccc} \frac{2}{27}&0&-\frac{\sqrt{2}}{27\sqrt{35}}&-\frac{2}{27\sqrt{7}} \\ 0&\frac{1}{45}&0&\frac{\sqrt{2}}{15\sqrt{7}}\\-\frac{\sqrt{2}}{27\sqrt{35}}&0&\frac{29}{189}&-\frac{5\sqrt{10}}{189}\\\frac{2}{27\sqrt{7}}
&\frac{\sqrt{2}}{15\sqrt{7}}&\frac{5\sqrt{10}}{189}&-\frac{28}{135}\end{array}\right)$\\
$\mathcal{A}_{18}[2]=\left(\begin{array}{cccc} 0&0&-\frac{\sqrt{7}}{54\sqrt{10}}&\frac{\sqrt{7}}{54}\\ 0&-\frac{7}{180}&0&-\frac{1}{15\sqrt{14}}\\-\frac{\sqrt{7}}{54\sqrt{10}}&0&-\frac{1}{252}&\frac{2\sqrt{10}}{189}\\-\frac{\sqrt{7}}{54}&-\frac{1}{15\sqrt{14}}&-\frac{2\sqrt{10}}{189}&\frac{1}{135}\end{array}\right)$ &$\mathcal{A}_{18}^{\prime}[2]=\left(\begin{array}{cccc} 0&0&-\frac{\sqrt{7}}{54\sqrt{10}}&-\frac{\sqrt{7}}{54}\\ 0&-\frac{7}{180}&0&-\frac{1}{15\sqrt{14}}\\-\frac{\sqrt{7}}{54\sqrt{10}}&0&-\frac{1}{252}&-\frac{2\sqrt{10}}{189}\\\frac{\sqrt{7}}{54}&-\frac{1}{15\sqrt{14}}&\frac{2\sqrt{10}}{189}&\frac{1}{135}\end{array}\right)$&$\begin{array}{l} \mathcal{A}_{13}^{(\prime)}[3]=\rm {diag}(1,1,1,1) \\ \mathcal{A}_{14}^{(\prime)}[3]=\rm {diag}(-1,\frac{3}{2},\frac{1}{2},-1)\\\mathcal{A}_{16}^{(\prime)}[3]=\rm {diag}(-\frac{1}{27},\frac{1}{18},\frac{5}{54},-\frac{1}{27})\\\mathcal{A}_{19}^{(\prime)}[3]=\rm {diag}(1,\frac{1}{6},\frac{1}{2},1)\end{array}$\\
$\mathcal{A}_{15}[3]=\left(\begin{array}{cccc} 0&\frac{3}{5\sqrt{2}}&-\frac{1}{\sqrt{5}}&-\frac{4\sqrt{3}}{5} \\ \frac{3}{5\sqrt{2}}&-\frac{3}{35}&-\frac{6\sqrt{2}}{7\sqrt{5}}&-\frac{6\sqrt{6}}{35}\\-\frac{1}{\sqrt{5}}&-\frac{6\sqrt{2}}{7\sqrt{5}}&-\frac{4}{7}&\frac{\sqrt{3}}{7\sqrt{5}}\\-\frac{4\sqrt{3}}{5}
&-\frac{6\sqrt{6}}{35}&\frac{\sqrt{3}}{7\sqrt{5}}&-\frac{22}{35}\end{array}\right)$&$\mathcal{A}_{15}^{\prime}[3]=\left(\begin{array}{cccc} 0&\frac{3}{5\sqrt{2}}&\frac{1}{\sqrt{5}}&-\frac{4\sqrt{3}}{5} \\ \frac{3}{5\sqrt{2}}&-\frac{3}{35}&\frac{6\sqrt{2}}{7\sqrt{5}}&-\frac{6\sqrt{6}}{35}\\\frac{1}{\sqrt{5}}&\frac{6\sqrt{2}}{7\sqrt{5}}&-\frac{4}{7}&-\frac{\sqrt{3}}{7\sqrt{5}}\\-\frac{4\sqrt{3}}{5}
&-\frac{6\sqrt{6}}{35}&-\frac{\sqrt{3}}{7\sqrt{5}}&-\frac{22}{35}\end{array}\right)$&$\mathcal{A}_{17}[3]=\left(\begin{array}{cccc} -\frac{4}{135}&\frac{\sqrt{2}}{105}&\frac{2\sqrt{5}}{189}&-\frac{4}{315\sqrt{3}} \\ \frac{\sqrt{2}}{105}&\frac{16}{315}&0&-\frac{\sqrt{2}}{35\sqrt{3}}\\-\frac{2\sqrt{5}}{189}&0&\frac{1}{21}&-\frac{\sqrt{5}}{63\sqrt{3}}\\-\frac{4}{315\sqrt{3}}
&-\frac{\sqrt{2}}{35\sqrt{3}}&\frac{\sqrt{5}}{63\sqrt{3}}&\frac{82}{945}\end{array}\right)$\\
$\mathcal{A}_{17}^{\prime}[3]=\left(\begin{array}{cccc} -\frac{4}{135}&\frac{\sqrt{2}}{105}&-\frac{2\sqrt{5}}{189}&-\frac{4}{315\sqrt{3}} \\ \frac{\sqrt{2}}{105}&\frac{16}{315}&0&-\frac{\sqrt{2}}{35\sqrt{3}}\\\frac{2\sqrt{5}}{189}&0&\frac{1}{21}&\frac{\sqrt{5}}{63\sqrt{3}}\\-\frac{4}{315\sqrt{3}}
&-\frac{\sqrt{2}}{35\sqrt{3}}&-\frac{\sqrt{5}}{63\sqrt{3}}&\frac{82}{945}\end{array}\right)$&$\mathcal{A}_{18}[3]=\left(\begin{array}{cccc} 0&\frac{1}{30\sqrt{2}}&\frac{\sqrt{5}}{54}&-\frac{1}{45\sqrt{3}}\\ \frac{1}{30\sqrt{2}}&\frac{1}{90}&0&-\frac{\sqrt{2}}{35\sqrt{3}}\\-\frac{\sqrt{5}}{54}&0&-\frac{2}{189}&\frac{\sqrt{5}}{126\sqrt{3}}\\-\frac{1}{45\sqrt{3}}&-\frac{\sqrt{2}}{35\sqrt{3}}&-\frac{\sqrt{5}}{126\sqrt{3}}&-\frac{11}{1890}\end{array}\right)$
&$\mathcal{A}_{18}^{\prime}[3]=\left(\begin{array}{cccc} 0&\frac{1}{30\sqrt{2}}&-\frac{\sqrt{5}}{54}&-\frac{1}{45\sqrt{3}}\\ \frac{1}{30\sqrt{2}}&\frac{1}{90}&0&-\frac{\sqrt{2}}{35\sqrt{3}}\\\frac{\sqrt{5}}{54}&0&-\frac{2}{189}&-\frac{\sqrt{5}}{126\sqrt{3}}\\-\frac{1}{45\sqrt{3}}&-\frac{\sqrt{2}}{35\sqrt{3}}&\frac{\sqrt{5}}{126\sqrt{3}}&-\frac{11}{1890}\end{array}\right)$      \\
\bottomrule[1.5pt]
\end{tabular}
\end{table*}


\begin{thebibliography}{99}
\bibitem{Choi:2003ue}
  S.~K.~Choi {\it et al.} [Belle Collaboration],
  Observation of a narrow charmonium-like state in exclusive $B^{\pm} \to K^{\pm} \pi^+ \pi^- J/\psi$ decays,
  \href{https://journals.aps.org/prl/abstract/10.1103/PhysRevLett.91.262001}{Phys.\ Rev.\ Lett.\  {\bf 91}, 262001 (2003)}.

\bibitem{Chen:2016qju}
  H.~X.~Chen, W.~Chen, X.~Liu, and S.~L.~Zhu,
  The hidden-charm pentaquark and tetraquark states,
  \href{http://linkinghub.elsevier.com/retrieve/pii/S037015731630103X}{Phys.\ Rep.\  {\bf 639}, 1 (2016)}.

\bibitem{Liu:2013waa}
  X.~Liu,
  An overview of $XYZ$ new particles,
  \href{http://dx.doi.org/10.1007/s11434-014-0407-2}{Chin.\ Sci.\ Bull.\  {\bf 59}, 3815 (2014)}.

\bibitem{Hosaka:2016pey}
  A.~Hosaka, T.~Iijima, K.~Miyabayashi, Y.~Sakai, and S.~Yasui,
  Exotic hadrons with heavy flavors: $X$, $Y$, $Z$, and related states,
  \href{http://dx.doi.org/10.1093/ptep/ptw045}{Prog. Theor. Exp. Phys. {\bf 2016}, 062C01 (2016)}.

\bibitem{Liu:2019zoy}
  Y.~R.~Liu, H.~X.~Chen, W.~Chen, X.~Liu and S.~L.~Zhu,
  Pentaquark and Tetraquark states,
  \href{https://www.sciencedirect.com/science/article/pii/S0146641019300304?via\%3Dihub}{Prog.\ Part.\ Nucl.\ Phys.\  {\bf 107}, 237 (2019)}.

\bibitem{Brambilla:2019esw}
N.~Brambilla, S.~Eidelman, C.~Hanhart, A.~Nefediev, C.~P.~Shen, C.~E.~Thomas, A.~Vairo and C.~Z.~Yuan,
The $XYZ$ states: experimental and theoretical status and perspectives,
\href{https://www.sciencedirect.com/science/article/pii/S0370157320301915?via\%3Dihub}{Phys. Rept. \textbf{873}, 1-154 (2020)}.

\bibitem{Olsen:2017bmm}
  S.~L.~Olsen, T.~Skwarnicki and D.~Zieminska,
  Nonstandard heavy mesons and baryons: Experimental evidence,
  \href{https://journals.aps.org/rmp/abstract/10.1103/RevModPhys.90.015003}{Rev.\ Mod.\ Phys.\  {\bf 90}, no. 1, 015003 (2018)}.

\bibitem{Guo:2017jvc}
  F.~K.~Guo, C.~Hanhart, U.~G.~Mei$\ss$ner, Q.~Wang, Q.~Zhao and B.~S.~Zou,
  Hadronic molecules,
  \href{https://journals.aps.org/rmp/abstract/10.1103/RevModPhys.90.015004}{Rev.\ Mod.\ Phys.\  {\bf 90}, no. 1, 015004 (2018)}.

\bibitem{Aaij:2019vzc}
  R.~Aaij {\it et al.} [LHCb Collaboration],
  Observation of a narrow pentaquark state, $P_c(4312)^+$, and of two-peak structure of the $P_c(4450)^+$,
 \href{https://journals.aps.org/prl/abstract/10.1103/PhysRevLett.122.222001}{Phys.\ Rev.\ Lett.\  {\bf 122}, no. 22, 222001 (2019)}.


\bibitem{Li:2014gra}
  X.~Q.~Li and X.~Liu,
  A possible global group structure for exotic states,
  \href{https://link.springer.com/article/10.1140\%2Fepjc\%2Fs10052-014-3198-3}{Eur.\ Phys.\ J.\ C {\bf 74}, no. 12, 3198 (2014)}.

\bibitem{Karliner:2015ina}
  M.~Karliner and J.~L.~Rosner,
  New Exotic Meson and Baryon Resonances from Doubly-Heavy Hadronic Molecules,
  \href{https://journals.aps.org/prl/abstract/10.1103/PhysRevLett.115.122001}{Phys.\ Rev.\ Lett.\  {\bf 115}, no. 12, 122001 (2015)}.

\bibitem{Wu:2010jy}
  J.~J.~Wu, R.~Molina, E.~Oset and B.~S.~Zou,
  Prediction of narrow $N^*$ and $\Lambda^*$ resonances with hidden charm above 4 GeV,
  \href{https://journals.aps.org/prl/abstract/10.1103/PhysRevLett.105.232001}{Phys.\ Rev.\ Lett.\  {\bf 105}, 232001 (2010)}.

\bibitem{Wang:2011rga}
  W.~L.~Wang, F.~Huang, Z.~Y.~Zhang and B.~S.~Zou,
  $\Sigma_c \bar{D}$ and $\Lambda_c \bar{D}$ states in a chiral quark model,
  \href{https://journals.aps.org/prc/abstract/10.1103/PhysRevC.84.015203}{Phys.\ Rev.\ C {\bf 84}, 015203 (2011)}.

\bibitem{Yang:2011wz}
  Z.~C.~Yang, Z.~F.~Sun, J.~He, X.~Liu and S.~L.~Zhu,
  The possible hidden-charm molecular baryons composed of anti-charmed meson and charmed baryon,
  \href{https://iopscience.iop.org/article/10.1088/1674-1137/36/1/002/meta}{Chin.\ Phys.\ C {\bf 36}, 6 (2012)}.

\bibitem{Wu:2012md}
  J.~J.~Wu, T.-S.~H.~Lee and B.~S.~Zou,
  Nucleon Resonances with Hidden Charm in Coupled-Channel Models,
  \href{https://journals.aps.org/prc/abstract/10.1103/PhysRevC.85.044002}{Phys.\ Rev.\ C {\bf 85}, 044002 (2012)}.

\bibitem{Chen:2015loa}
  R.~Chen, X.~Liu, X.~Q.~Li and S.~L.~Zhu,
  Identifying exotic hidden-charm pentaquarks,
  \href{https://journals.aps.org/prl/abstract/10.1103/PhysRevLett.115.132002}{Phys.\ Rev.\ Lett.\  {\bf 115}, 132002 (2015)}.


\bibitem{Wong:2003xk}
  C.~Y.~Wong,
  Molecular states of heavy quark mesons,
  \href{https://journals.aps.org/prc/abstract/10.1103/PhysRevC.69.055202}{Phys.\ Rev.\ C {\bf 69}, 055202 (2004)}.


\bibitem{Swanson:2003tb}
  E.~S.~Swanson,
  Short range structure in the $X(3872)$,
  \href{https://www.sciencedirect.com/science/article/pii/S0370269304004599?via\%3Dihub}{Phys.\ Lett.\ B {\bf 588}, 189 (2004)}.

\bibitem{Suzuki:2005ha}
  M.~Suzuki,
  The $X(3872)$ boson: Molecule or charmonium,
  \href{https://journals.aps.org/prd/abstract/10.1103/PhysRevD.72.114013}{Phys.\ Rev.\ D {\bf 72}, 114013 (2005)}.

\bibitem{Liu:2008fh}
  Y.~R.~Liu, X.~Liu, W.~Z.~Deng and S.~L.~Zhu,
  Is $X(3872)$ Really a Molecular State?,
  \href{https://link.springer.com/article/10.1140/epjc/s10052-008-0640-4}{Eur.\ Phys.\ J.\ C {\bf 56}, 63 (2008)}.

\bibitem{Thomas:2008ja}
  C.~E.~Thomas and F.~E.~Close,
  Is $X(3872)$ a molecule?,
  \href{https://journals.aps.org/prd/abstract/10.1103/PhysRevD.78.034007}{Phys.\ Rev.\ D {\bf 78}, 034007 (2008)}.

\bibitem{Liu:2008tn}
  X.~Liu, Z.~G.~Luo, Y.~R.~Liu and S.~L.~Zhu,
  $X(3872)$ and Other Possible Heavy Molecular States,
  \href{https://link.springer.com/article/10.1140\%2Fepjc\%2Fs10052-009-1020-4}{Eur.\ Phys.\ J.\ C {\bf 61}, 411 (2009)}.

\bibitem{Lee:2009hy}
  I.~W.~Lee, A.~Faessler, T.~Gutsche and V.~E.~Lyubovitskij,
  $X(3872)$ as a molecular $DD^*$ state in a potential model,
  \href{https://journals.aps.org/prd/abstract/10.1103/PhysRevD.80.094005}{Phys.\ Rev.\ D {\bf 80}, 094005 (2009)}.

\bibitem{Zhao:2014gqa}
L.~Zhao, L.~Ma and S.~L.~Zhu,
Spin-orbit force, recoil corrections, and possible $B \bar{B}^{*}$ and $D \bar{D}^{*}$  molecular states,
\href{https://journals.aps.org/prd/abstract/10.1103/PhysRevD.89.094026}{Phys. Rev. D \textbf{89}, no.9, 094026 (2014)}.

\bibitem{Li:2012cs}
N.~Li and S.~L.~Zhu,
Isospin breaking, Coupled-channel effects and Diagnosis of $X(3872)$,
\href{https://journals.aps.org/prd/abstract/10.1103/PhysRevD.86.074022}{Phys. Rev. D \textbf{86}, 074022 (2012)}.

\bibitem{Choi:2007wga}
  S.~K.~Choi {\it et al.} [Belle Collaboration],
  Observation of a resonance-like structure in the $\pi^\pm \psi^\prime$ mass distribution in exclusive $B \to K \pi^\pm \psi^\prime$ decays,
  \href{https://journals.aps.org/prl/abstract/10.1103/PhysRevLett.100.142001}{Phys.\ Rev.\ Lett.\  {\bf 100}, 142001 (2008)}.

\bibitem{Liu:2007bf}
  X.~Liu, Y.~R.~Liu, W.~Z.~Deng and S.~L.~Zhu,
  Is $Z^+(4430)$ a loosely bound molecular state?,
  \href{https://journals.aps.org/prd/abstract/10.1103/PhysRevD.77.034003}{Phys.\ Rev.\ D {\bf 77}, 034003 (2008)}.

\bibitem{Liu:2008xz}
  X.~Liu, Y.~R.~Liu, W.~Z.~Deng and S.~L.~Zhu,
  $Z^+(4430)$ as a $D_1^\prime D^* (D_1 D^*)$ molecular state,
  \href{https://journals.aps.org/prd/abstract/10.1103/PhysRevD.77.094015}{Phys.\ Rev.\ D {\bf 77}, 094015 (2008)}.

\bibitem{Close:2009ag}
  F.~Close and C.~Downum,
  On the possibility of Deeply Bound Hadronic Molecules from single Pion Exchange,
  \href{https://journals.aps.org/prl/abstract/10.1103/PhysRevLett.102.242003}{Phys.\ Rev.\ Lett.\  {\bf 102}, 242003 (2009)}.

\bibitem{Aaltonen:2009tz}
  T.~Aaltonen {\it et al.} [CDF Collaboration],
  Evidence for a Narrow Near-Threshold Structure in the $J/\psi\phi$ Mass Spectrum in $B^+\to J/\psi\phi K^+$ Decays,
  \href{https://journals.aps.org/prl/abstract/10.1103/PhysRevLett.102.242002}{Phys.\ Rev.\ Lett.\  {\bf 102}, 242002 (2009)}.

\bibitem{Liu:2009ei}
X.~Liu and S.~L.~Zhu,
$Y(4143)$ is probably a molecular partner of $Y(3930)$,
\href{https://journals.aps.org/prd/abstract/10.1103/PhysRevD.80.017502}{Phys. Rev. D \textbf{80}, 017502 (2009)}.

\bibitem{Uehara:2005qd}
  S.~Uehara {\it et al.} [Belle Collaboration],
  Observation of a $\chi_{c2}^\prime$ candidate in $\gamma \gamma \to D \bar D$ production at BELLE,
  \href{https://journals.aps.org/prl/abstract/10.1103/PhysRevLett.96.082003}{Phys.\ Rev.\ Lett.\  {\bf 96}, 082003 (2006)}.

\bibitem{Aubert:2005rm}
  B.~Aubert {\it et al.} [BaBar Collaboration],
  Observation of a broad structure in the $\pi^+ \pi^- J/\psi$ mass spectrum around 4.26-GeV/c$^2$,
  \href{https://journals.aps.org/prl/abstract/10.1103/PhysRevLett.95.142001}{Phys.\ Rev.\ Lett.\  {\bf 95}, 142001 (2005)}.

\bibitem{Ding:2008gr}
  G.~J.~Ding,
  Are $Y(4260)$ and {\rm$Z_2^{+}$(4250)} ${\rm D_1D}$ or ${\rm D_0D^{*}}$ hadronic molecules?
  \href{https://journals.aps.org/prd/abstract/10.1103/PhysRevD.79.014001}{Phys.\ Rev.\ D {\bf 79}, 014001 (2009)}.

\bibitem{Cleven:2013mka}
  M.~Cleven, Q.~Wang, F.~K.~Guo, C.~Hanhart, U.~G.~Mei$\beta$ner and Q.~Zhao,
  $Y(4260)$ as the first $S$-wave open charm vector molecular state?,
  \href{https://journals.aps.org/prd/abstract/10.1103/PhysRevD.90.074039}{Phys.\ Rev.\ D {\bf 90}, no. 7, 074039 (2014)}.


\bibitem{Wang:2013kra}
  Q.~Wang, M.~Cleven, F.~K.~Guo, C.~Hanhart, U.~G.~Mei$\beta$ner, X.~G.~Wu and Q.~Zhao,
  $Y(4260)$: hadronic molecule versus hadro-charmonium interpretation,
  \href{https://journals.aps.org/prd/abstract/10.1103/PhysRevD.89.034001}{Phys.\ Rev.\ D {\bf 89}, no. 3, 034001 (2014)}.

\bibitem{Sun:2012zzd}
  Z.~F.~Sun, Z.~G.~Luo, J.~He, X.~Liu and S.~L.~Zhu,
  A note on the $B^* \bar B$, $B^* \bar B$, $D^* \bar D$, $D^* \bar D$ molecular states,
  \href{https://iopscience.iop.org/article/10.1088/1674-1137/36/3/002}{Chin.\ Phys.\ C {\bf 36}, 194 (2012)}.

\bibitem{Sun:2012sy}
Z.~F.~Sun, X.~Liu, M.~Nielsen and S.~L.~Zhu,
Hadronic molecules with both open charm and bottom,
\href{https://journals.aps.org/prd/abstract/10.1103/PhysRevD.85.094008}{Phys. Rev. D \textbf{85}, 094008 (2012)}.

\bibitem{Hu:2010fg}
B.~Hu, X.~L.~Chen, Z.~G.~Luo, P.~Z.~Huang, S.~L.~Zhu, P.~F.~Yu and X.~Liu,
Possible heavy molecular states composed of a pair of excited charm-strange mesons,
\href{https://iopscience.iop.org/article/10.1088/1674-1137/35/2/002}{Chin. Phys. C \textbf{35}, 113-125 (2011)}.

\bibitem{Shen:2010ky}
L.~L.~Shen, X.~L.~Chen, Z.~G.~Luo, P.~Z.~Huang, S.~L.~Zhu, P.~F.~Yu and X.~Liu,
The Molecular systems composed of the charmed mesons in the $H\bar{S}+h.c.$ doublet,
\href{https://link.springer.com/article/10.1140/epjc/s10052-010-1441-0}{Eur. Phys. J. C \textbf{70}, 183-217 (2010)}.

\bibitem{Chen:2015add}
R.~Chen, X.~Liu, Y.~R.~Liu and S.~L.~Zhu,
Predictions of the hidden-charm molecular states with four-quark component,
\href{https://link.springer.com/article/10.1140/epjc/s10052-016-4166-x}{Eur. Phys. J. C \textbf{76}, no.6, 319 (2016)}.

\bibitem{Aaij:2016nsc}
  R.~Aaij {\it et al.} [LHCb Collaboration],
  Amplitude analysis of $B^+\to J/\psi \phi K^+$ decays,
  \href{https://journals.aps.org/prd/abstract/10.1103/PhysRevD.95.012002}{Phys.\ Rev.\ D {\bf 95}, no. 1, 012002 (2017)}.

\bibitem{Liu:2010hf}
  X.~Liu, Z.~G.~Luo and S.~L.~Zhu,
  Novel charmonium-like structures in the $J/\psi\phi$ and $J/\psi\omega$ invariant mass spectra,
  \href{https://www.sciencedirect.com/science/article/pii/S0370269311014766?via\%3Dihub}{Phys.\ Lett.\ B {\bf 699}, 341 (2011)}
  Erratum: \href{https://www.sciencedirect.com/science/article/pii/S0370269311004230?via\%3Dihub}{[Phys.\ Lett.\ B {\bf 707}, 577 (2012)]}

\bibitem{Aaij:2016iza}
  R.~Aaij {\it et al.} [LHCb Collaboration],
  Observation of $J/\psi\phi$ structures consistent with exotic states from amplitude analysis of $B^+\to J/\psi \phi K^+$ decays,
  \href{https://journals.aps.org/prl/abstract/10.1103/PhysRevLett.118.022003}{Phys.\ Rev.\ Lett.\  {\bf 118}, no. 2, 022003 (2017)}.

\bibitem{Abazov:2007sf}
  V.~M.~Abazov {\it et al.} [D0 Collaboration],
 Measurement of the $\Lambda_b$ Lifetime in the Exclusive Decay $\Lambda_b \to J/\psi \Lambda$,
  \href{https://journals.aps.org/prl/abstract/10.1103/PhysRevLett.99.142001}{Phys.\ Rev.\ Lett.\  {\bf 99}, 142001 (2007)}.


\bibitem{Jia:2019gfe}
  S.~Jia {\it et al.} [Belle Collaboration],
  Observation of a vector charmoniumlike state in $e^+e^- \to D^+_sD_{s1}(2536)^-+c.c.$,
  \href{https://journals.aps.org/prd/abstract/10.1103/PhysRevD.100.111103}{Phys.\ Rev.\ D {\bf 100}, no. 11, 111103 (2019)}.


\bibitem{Pakhlova:2008vn}
  G.~Pakhlova {\it et al.} [Belle Collaboration],
 Observation of a near-threshold enhancement in the $e^+e^- \to \Lambda_c^+ \Lambda_c^-$ cross section using initial-state radiation,
  \href{https://journals.aps.org/prl/abstract/10.1103/PhysRevLett.101.172001}{Phys.\ Rev.\ Lett.\  {\bf 101}, 172001 (2008)}.

\bibitem{Ablikim:2019hff}
M.~Ablikim \textit{et al.} [BESIII],
Future Physics Programme of BESIII,
\href{https://iopscience.iop.org/article/10.1088/1674-1137/44/4/040001}{Chin. Phys. C \textbf{44}, no.4, 040001 (2020)}.

\bibitem{Wang:2019mhs}
  J.~Z.~Wang, D.~Y.~Chen, X.~Liu and T.~Matsuki,
  Constructing $J/\psi$ family with updated data of charmoniumlike $Y$ states,
  \href{https://journals.aps.org/prd/abstract/10.1103/PhysRevD.99.114003}{Phys.\ Rev.\ D {\bf 99}, no. 11, 114003 (2019)}.

\bibitem{Wang:2020prx}
  J.~Z.~Wang, R.~Q.~Qian, X.~Liu and T.~Matsuki,
  Are the $Y$ states around 4.6 GeV from $e^+e^-$ annihilation higher charmonia?,
  \href{https://journals.aps.org/prd/abstract/10.1103/PhysRevD.101.034001}{Phys.\ Rev.\ D {\bf 101}, no. 3, 034001 (2020)}.

\bibitem{Tornqvist:1993ng}
  N.~A.~Tornqvist,
  From the deuteron to deusons, an analysis of deuteron-like meson-meson bound states,
   \href{https://link.springer.com/article/10.1007\%2FBF01413192}{Z.\ Phys.\ C {\bf 61}, 525 (1994)}.

\bibitem{Tornqvist:1993vu}
  N.~A.~Tornqvist,
  On deusons or deuteron-like meson-meson bound states,
  \href{https://link.springer.com/article/10.1007\%2FBF02734018}{Nuovo Cimento.\ A {\bf 107}, 2471 (1994)}.

\bibitem{Wise:1992hn}
  M.~B.~Wise,
  Chiral perturbation theory for hadrons containing a heavy quark,
   \href{https://journals.aps.org/prd/abstract/10.1103/PhysRevD.45.R2188}{Phys.\ Rev.\ D {\bf 45}, R2188 (1992)}.

\bibitem{Casalbuoni:1992gi}
  R.~Casalbuoni, A.~Deandrea, N.~Di Bartolomeo, R.~Gatto, F.~Feruglio, and G.~Nardulli,
  Light vector resonances in the effective chiral Lagrangian for heavy mesons,
  \href{https://www.sciencedirect.com/science/article/abs/pii/037026939291189G}{Phys.\ Lett.\ B {\bf 292}, 371 (1992)}.

\bibitem{Casalbuoni:1996pg}
  R.~Casalbuoni, A.~Deandrea, N.~Di Bartolomeo, R.~Gatto, F.~Feruglio, and G.~Nardulli,
  Phenomenology of heavy meson chiral Lagrangians,
  \href{https://www.sciencedirect.com/science/article/pii/S0370157396000270}{Phys.\ Rep.\  {\bf 281}, 145 (1997)}.

\bibitem{Yan:1992gz}
  T.~M.~Yan, H.~Y.~Cheng, C.~Y.~Cheung, G.~L.~Lin, Y.~C.~Lin, and H.~L.~Yu,
  Heavy quark symmetry and chiral dynamics,
  \href{https://journals.aps.org/prd/abstract/10.1103/PhysRevD.46.1148}{Phys.\ Rev.\ D {\bf 46}, 1148 (1992);}
   \href{https://journals.aps.org/prd/abstract/10.1103/PhysRevD.55.5851}{[Phys.\ Rev.\ D {\bf 55}, 5851E (1997)]}.

\bibitem{Cheng:2010yd}
  H.~Y.~Cheng and K.~C.~Yang,
  Charmless hadronic $B$ decays into a tensor meson,
  \href{https://journals.aps.org/prd/abstract/10.1103/PhysRevD.83.034001}{Phys.\ Rev.\ D {\bf 83}, 034001 (2011)}.

\bibitem{Wang:2019nwt}
  F.~L.~Wang, R.~Chen, Z.~W.~Liu and X.~Liu,
  Probing new types of $P_c$ states inspired by the interaction between $S$-wave charmed baryon and anti-charmed meson in a $\bar T$ doublet,
  \href{https://journals.aps.org/prc/abstract/10.1103/PhysRevC.101.025201}{Phys.\ Rev.\ C {\bf 101}, no. 2, 025201 (2020)}.

\bibitem{Breit:1929zz}
  G.~Breit,
  The Effect of Retardation on the Interaction of Two Electrons,
  \href{https://journals.aps.org/pr/abstract/10.1103/PhysRev.34.553}{Phys.\ Rev.\  {\bf 34}, 553 (1929)}.

\bibitem{Breit:1930zza}
  G.~Breit,
  The Fine Structure of HE as a Test of the Spin Interactions of Two Electrons,
  \href{https://journals.aps.org/pr/abstract/10.1103/PhysRev.36.383}{Phys.\ Rev.\  {\bf 36}, 383 (1930)}.

\bibitem{Klempt:2002ap}
  E.~Klempt, F.~Bradamante, A.~Martin, and J.~M.~Richard,
  Antinucleon nucleon interaction at low energy: Scattering and protonium,
  \href{http://www.sciencedirect.com/science/article/pii/S0370157302001448?via\%3Dihub}{Phys.\ Rep.\  {\bf 368}, 119 (2002)}.

\bibitem{Falk:1992cx}
A.~F.~Falk and M.~E.~Luke, Strong decays of excited heavy mesons in chiral perturbation theory,
\href{https://www.sciencedirect.com/science/article/abs/pii/037026939290618E?via\%3Dihub}{Phys.\ Lett.\  B {\bf 292}, 119 (1992)}.

\bibitem{Isola:2003fh}
  C.~Isola, M.~Ladisa, G.~Nardulli, and P.~Santorelli,
  Charming penguins in $B\to K^{*}\pi, K(\rho,\omega,\phi)$ decays,
  \href{https://journals.aps.org/prd/abstract/10.1103/PhysRevD.68.114001}{Phys.\ Rev.\ D {\bf 68}, 114001 (2003)}.

\bibitem{Cleven:2016qbn}
  M.~Cleven and Q.~Zhao,
  Cross section line shape of $e^+e^-\to\chi_{c0}\omega$ around the $Y(4260)$ mass region,
 \href{https://linkinghub.elsevier.com/retrieve/pii/S0370269317301545}{Phys.\ Lett.\ B {\bf 768}, 52 (2017)}.

\bibitem{Dong:2019ofp}
X.~K.~Dong, Y.~H.~Lin and B.~S.~Zou,
Prediction of an exotic state around 4240 MeV with $J^{PC}=1^{-+}$ as C-parity partner of $Y(4260)$ in molecular picture,
\href{https://journals.aps.org/prd/abstract/10.1103/PhysRevD.101.076003}{Phys. Rev. D \textbf{101}, no.7, 076003 (2020)}.

\bibitem{He:2019csk}
J.~He, Y.~Liu, J.~T.~Zhu and D.~Y.~Chen,
$Y(4626)$ as a molecular state from interaction ${D}^*_s{\bar{D}}_{s1}(2536)-{D}_s{\bar{D}}_{s1}(2536)$,
\href{https://link.springer.com/article/10.1140/epjc/s10052-020-7820-2}{Eur. Phys. J. C \textbf{80}, no.3, 246 (2020)}.

\bibitem{Wang:2019aoc}
  F.~L.~Wang, R.~Chen, Z.~W.~Liu and X.~Liu,
  Possible triple-charm molecular pentaquarks from $\Xi_{cc}D_1/\Xi_{cc}D_2^*$ interactions,
  \href{https://journals.aps.org/prd/abstract/10.1103/PhysRevD.99.054021}{Phys.\ Rev.\ D {\bf 99}, 054021 (2019)}.

\bibitem{Wang:2020lua}
Z.~Y.~Wang, J.~J.~Qi, J.~Xu and X.~H.~Guo,
Studying the $D_1D$ molecule in the Bethe-Salpeter equation approach,
\href{https://journals.aps.org/prd/abstract/10.1103/PhysRevD.102.036008}{Phys. Rev. D \textbf{102}, no.3, 036008 (2020)}.

\bibitem{Riska:2000gd}
  D.~O.~Riska and G.~E.~Brown,
  Nucleon resonance transition couplings to vector mesons,
  \href{https://www.sciencedirect.com/science/article/pii/S0375947400003626}{Nucl.\ Phys.\ {\bf A} {\bf 679}, 577 (2001)}.

\bibitem{Tanabashi:2018oca}
  M.~Tanabashi {\it et al.} (Particle Data Group),
  Review of particle physics,
  \href{https://journals.aps.org/prd/abstract/10.1103/PhysRevD.98.030001}{Phys.\ Rev.\ D {\bf 98}, 030001 (2018)}.

\bibitem{Chen:2017xat}
  R.~Chen, A.~Hosaka and X.~Liu,
  Searching for possible $\Omega_c$-like molecular states from meson-baryon interaction,
  \href{https://journals.aps.org/prd/abstract/10.1103/PhysRevD.97.036016}{Phys.\ Rev.\ D {\bf 97}, 036016 (2018)}.

\end{thebibliography}
\end{document}